\documentclass[fleqn,usenatbib]{mnras} %Class file altered on lines 116, 119, 122 and 123.
\usepackage[T1]{fontenc}
\usepackage{ae,aecompl}
\usepackage{newtxtext,newtxmath}

\usepackage{graphicx}	% Including figure files
\usepackage{amsmath}	% Advanced maths commands
\usepackage{amssymb}	% Extra maths symbols
\usepackage{amsmath}
\usepackage{wasysym}
\usepackage{cite}

\newcommand{\appropto}{\mathrel{\vcenter{
		\offinterlineskip\halign{\hfil$##$\cr %The $##$ is not a mistake!
			\propto\cr\noalign{\kern2pt}\sim\cr\noalign{\kern-2pt}}}}}

\title[NGC 1052-DF2 verifies MOND predictions]{A new formulation of the external field effect in MOND and numerical simulations of ultra-diffuse dwarf galaxies $-$ application to {NGC 1052-DF2} and {NGC 1052-DF4}}

\author[H. Haghi et al.]{\parbox[t]{\textwidth}{Hosein Haghi$^{1}$\thanks{E-mail: haghi@iasbs.ac.ir}, Pavel~Kroupa $^{2, 3}$, Indranil Banik$^{2, 4}$, Xufen Wu$^{5, 6}$, Akram Hasani Zonoozi$^{1, 2}$, Behnam Javanmardi$^{7, 8}$, Amir Ghari$^{1}$, Oliver M\"uller$^{9}$, J{\"o}rg Dabringhausen$^{2}$ and Hongsheng Zhao$^{4, 10}$}\vspace{10pt}
\\
$^{1}$Institute for Advanced Studies in Basic Sciences, Physics department, Gavazang road, Zanjan, 4513766731 Zanjan, Iran \\
$^{2}$Helmholtz-Institut f\"ur Strahlen- und Kernphysik, University of Bonn, Nussallee 14-16, D-53115 Bonn, Germany\\
$^{3}$Charles University, Faculty of Mathematics and Physics, Astronomical Institute, V Hole\v{s}ovi\v{c}k\'ach 2, CZ-18000 Praha 8, Czech Republic\\
$^{4}$Scottish Universities Physics Alliance, University of St Andrews, North Haugh, St Andrews, Fife KY16 9SS, UK\\
$^{5}$CAS Key Laboratory for Research in Galaxies and Cosmology, Department of Astronomy, University of Science and Technology of China,\\ Hefei, 230026, P.R. China\\
$^{6}$School of Astronomy and Space Science, University of Science and Technology of China, Hefei 230026, China\\
$^{7}$School of Astronomy, Institute for Research in Fundamental Sciences (IPM) P. O. Box 19395-5531, Tehran, Iran\\
$^{8}$LESIA, Observatoire de Paris, Univ. PSL, CNRS, Sorbonne Univ., Univ. Paris Diderot, Sorbonne Paris Cit\'e, 5 Place Jules Janssen, 92195 Meudon, France \\
$^{9}$Observatoire Astronomique de Strasbourg  (ObAS), Universite de Strasbourg - CNRS, UMR 7550 Strasbourg, France\\
$^{10}$Department of Physics and Astronomy, LaserLaB, Vrije Universiteit, De Boelelaan 1081, NL-1081 HV Amsterdam, the Netherlands}

\pubyear{2019}
\begin{document}
\label{firstpage}
\pagerange{\pageref{firstpage}--\pageref{lastpage}}
\maketitle

\begin{abstract}

The ultra-diffuse dwarf galaxy NGC 1052-DF2 (DF2) has ten (eleven) measured globular clusters (GCs) with a line-of-sight velocity dispersion of $\sigma=7.8^{+5.2}_{-2.2}\,$km/s ($\sigma=10.6^{+3.9}_{-2.3}\,$km/s). Our conventional statistical analysis of the original ten GCs gives $\sigma=8.0^{+4.3}_{-3.0}\,$km/s. The overall distribution of velocities agrees well with a Gaussian of this width. Due to the non-linear Poisson equation in MOND, a dwarf galaxy has weaker self-gravity when in close proximity to a massive host. This external field effect is investigated using a new analytic formulation and fully self-consistent live $N$-body models in MOND. Our formulation agrees well with that of Famaey and McGaugh (2012). These new simulations confirm our analytic results and suggest that DF2 may be in a deep-freeze state unique to MOND. The correctly calculated MOND velocity dispersion agrees with our inferred dispersion and that of van Dokkum et al. (2018b) if DF2 is within 150 kpc of NGC 1052 and both are 20 Mpc away. The GCs of DF2 are however significantly brighter and larger than normal GCs, a problem which disappears if DF2 is significantly closer to us. A distance of 10-13 Mpc makes DF2 a normal dwarf galaxy even more consistent with MOND and the 13 Mpc distance reported by Trujillo et. al. (2019). We discuss the similar dwarf DF4, finding good agreement with MOND. We also discuss possible massive galaxies near DF2 and DF4 along with their distances and peculiar velocities, noting that NGC 1052 may lie at a distance near 10 Mpc.

%The possible host galaxies to which DF2 and DF4 may be satellites and their distances and peculiar velocities are also discussed, noting that a distance near 10 Mpc of NGC 1052 is not excluded.
%This distance is implied using the SBF method if DF2 has a stellar population consistent with that of other low-surface brightness galaxies as explainable by the IGIMF theory.
%IB: word limit of 250 reached, any further changes must not add any words overall.
%but would challenge the 20 Mpc distance reported by van Dokkum et al. (2018). It may also hint at non-canonical stellar populations in these unusual dwarf galaxies.
% We note that both may be normal satellites of the disk galaxy NGC 1042.
%IB_OM: Last sentence deleted.

\end{abstract}

\begin{keywords}
	gravitation -- dark matter -- galaxies: dwarf -- galaxies: kinematics and dynamics -- galaxies: distances and redshifts -- galaxies: individual: NGC 1052-DF2
\end{keywords}

\section{Introduction}

Amongst the most competitive solutions to the missing mass problem are the standard cosmological $\Lambda$CDM model \citep{Ostriker_Steinhardt_1995} and the Milgromian dynamics (MOND) theory, which was proposed by \citet{Milgrom83} at a similar time to when the notion of dark matter came to be taken seriously \citep{OP73}. Although it is generally thought that the dark matter model is successful on large scales \citep[e.g.][]{Planck16}, dark matter particles have not been detected after much experimental effort \citep[e.g.][]{Hoof_2019}. Moreover, the results of high-resolution $N$-body simulations do not seem to be compatible with observations on galactic and cosmological scales \citep{Haslbauer_2019, Bose_2018, Bullock_2017, PN10, Kroupa10, Kroupa12a, Kroupa15}.

MOND can be formulated as space-time scale invariance \citep{Milgrom09,WK15}. This is an excellent description of gravitation within Milgromian dynamics \citep{Milgrom83, BM84, FM12, Bullock_2017}. In MOND, a galaxy with an internal acceleration larger than Milgrom's constant $a_0\approx 3.8\,$pc/Myr$^2$; is in the Newtonian-gravitational regime which breaks space-time scale invariance, while for lower accelerations the equations of motion are space-time scale invariant, representing the MOND regime. MOND predicts that each isolated galaxy has a phantom dark matter halo which can be described mathematically as a Newtonian isothermal potential, causing the gravitating mass of the galaxy to exceed its inertial mass composed of normal, baryonic matter. In the external field of another galaxy, the non-linear MOND theory predicts this phantom dark matter halo to be reduced such that the internal dynamics of a system depends on the positions of nearby galaxies, even if they exert no tides. This external field effect (EFE) constitutes an important prediction of MOND which follows directly from its governing equations \citep{BM84, Milgrom1986}.

Observationally, the EFE can be tested by studying low-mass dwarf galaxies in the vicinity of major host galaxies. It has been successfully applied in the Local Group \citep{MM13}, in particular to correctly predict the very low velocity dispersion of Crater 2 \citep{McGaugh16a, Caldwell_2017}. Evidence for the EFE has been found in the rotation curves of galaxies \citep{WK15, Hees16, Haghi16}, the Milky Way escape velocity curve \citep{banik18} and in the asymmetric tidal tail of the globular star cluster Pal~5 \citep{Thomas+18}.

In this context, it is interesting to note the observations by \citet{vanDokkum+18c} of NGC 1052-Dragonfly 2 (DF2)\footnote{For a review of the discovery history and proper name of this galaxy (here referred to in short as DF2), see \citet{Trujillo2018}}, which was previously discovered by \citet[][plate 1]{Fosbury_1978}. \citet{vanDokkum+18c} used the line-of-sight velocities of its 10 GCs as bright tracers of its potential, which is consistent without dark matter in a Newtonian context.

Another dwarf galaxy was recently discovered by \citet{vD19_DF4}. NGC 1052-DF4 (DF4) is in close projected proximity to DF2 with similar unusual size, luminosity, morphology, globular cluster population and velocity dispersion. Based on the radial velocities of 7 GCs associated with DF4, they derived a Newtonian dynamical mass-to-light ratio of about unity.

These galaxies are studied here in order to test MOND and shed additional light on their possible origin. In Section \ref{sec:analytic}, we introduce a new set of fitting functions to calculate the global line-of-sight velocity dispersion of a non-isolated stellar system lying in the external field of a host galaxy as a function of the internal and external gravitational field. Our detailed formulation is compared with the EFE formulation by \citet{FM12}. In Section \ref{sec:veldisp}, we compare the MOND-predicted velocity dispersion of the GC system with the observed velocity dispersion of DF2. The first $N$-body numerical MOND modelling of DF2 is also documented in Section \ref{sec:Nbody}. We then address the unusual appearance of DF2, finding that if it is only 13 Mpc away instead of the 20 Mpc distance estimated by \citet{vanDokkum+18c}, it would also be consistent with MOND, even if it were isolated (Section \ref{sec:Distance}). A smaller distance of 13 Mpc was in fact recently suggested by \citet{Trujillo2018}. In Section \ref{sec:DF4}, we apply our analytical formalism to the recently discovered NGC 1052-DF4 \citep[DF4,][]{vD19_DF4}. Our results show that it can also be explained in MOND thanks to its weak self-gravity, which renders it susceptible to the EFE. We provide our conclusions in Section \ref{sec:conclusion}, emphasising that, given current measurement uncertainties, the NGC 1052, DF2 and DF4 system may be at a distance of $\approx 10$ Mpc.

\section{Analytic description of the velocity dispersion} \label{sec:analytic}

The velocity dispersion $\sigma$ of DF2 is a measure of its potential assuming virial equilibrium. In this section, we obtain the MOND expectation for $\sigma$. This prediction has no free parameters but depends on the dwarf's baryonic mass $M_{\rm DF2}$, effective radius $r_e$,  distance $D$ from the observer and its separation $D_{\rm sep}$ from the host galaxy with baryonic mass $M_{\rm NGC1052}$, which defines the external field.

A formulation of how the velocity dispersion of a self-gravitating system depends on the internal and external fields is made available here in the form of analytical functions. This formulation is equivalent to but generalizes that available in \citet{Kroupa_2018}. If DF2 is in dynamical equilibrium at a separation $D_{\rm sep}$ from NGC 1052, then the line-of-sight velocity dispersion $\sigma_{\rm M, EF}$ can be calculated explicitly as a function of the internal and external accelerations in MOND.

%Text below altered slightly by IB.
The globally averaged one-dimensional line-of-sight velocity dispersion $\sigma_{\rm M, EF}$ of a non-isolated stellar system, when the external gravity is much weaker or stronger than the internal gravity, and also in the intermediate regime was quantified by \citet{Haghi09}, using the numerical MOND potential solver code \textsc{N-MODY} \citep{NMODY}. \citet{Haghi09} formulated a functional representation for $\sigma_{\rm M, EF}$ in the intermediate regime ($a_i \approx a_e \leq a_0$) for different values of the external field and quantified the different asymptotic behaviour (i.e. in the Newtonian, the deep-MOND and the external-field-dominated regimes). The formulation was presented as their equations 16 and 17 with coefficients provided for different values of $a_e$ in their table 1. Here, an analytical formulation is found for the data in that table, allowing $\sigma_{\rm M, EF}$ to be calculated as a function of the internal field in a system with mass $M$ exposed to an external field $a_e$.

\begin{figure*}
	\centering
	\includegraphics[width = 5.6cm] {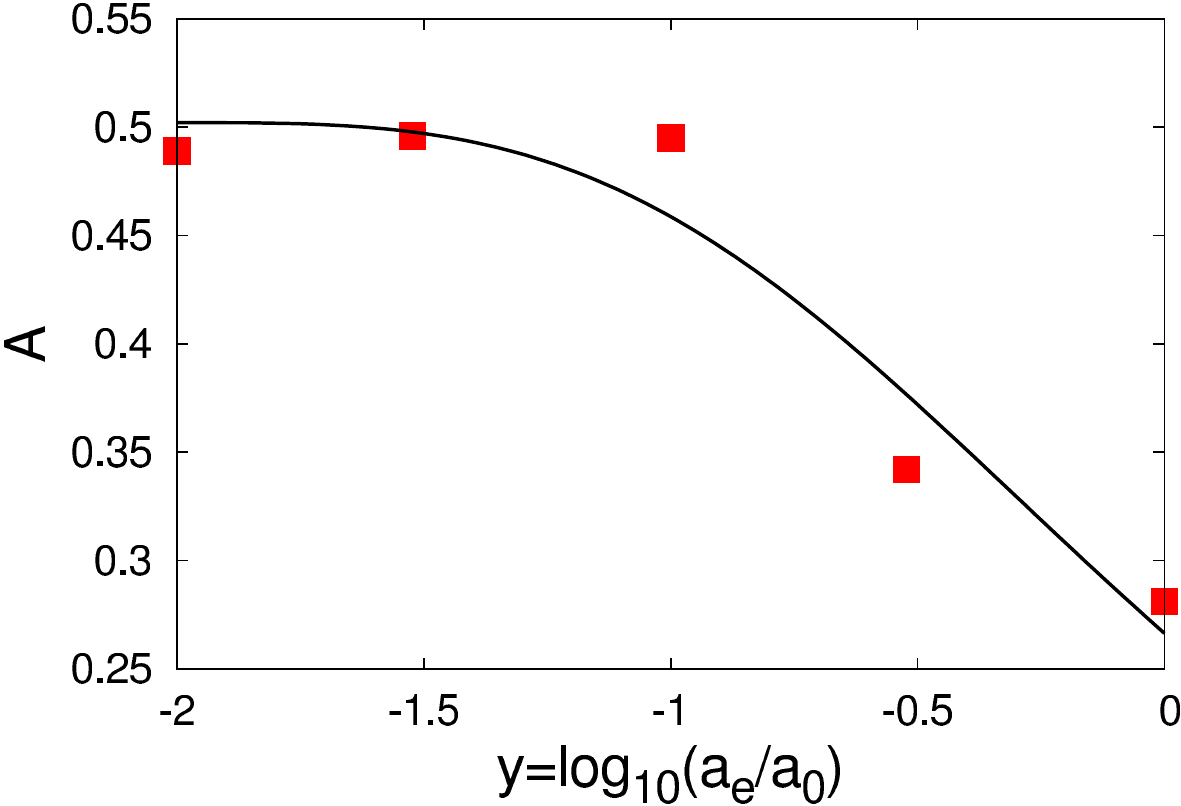}
	\includegraphics[width = 5.6cm] {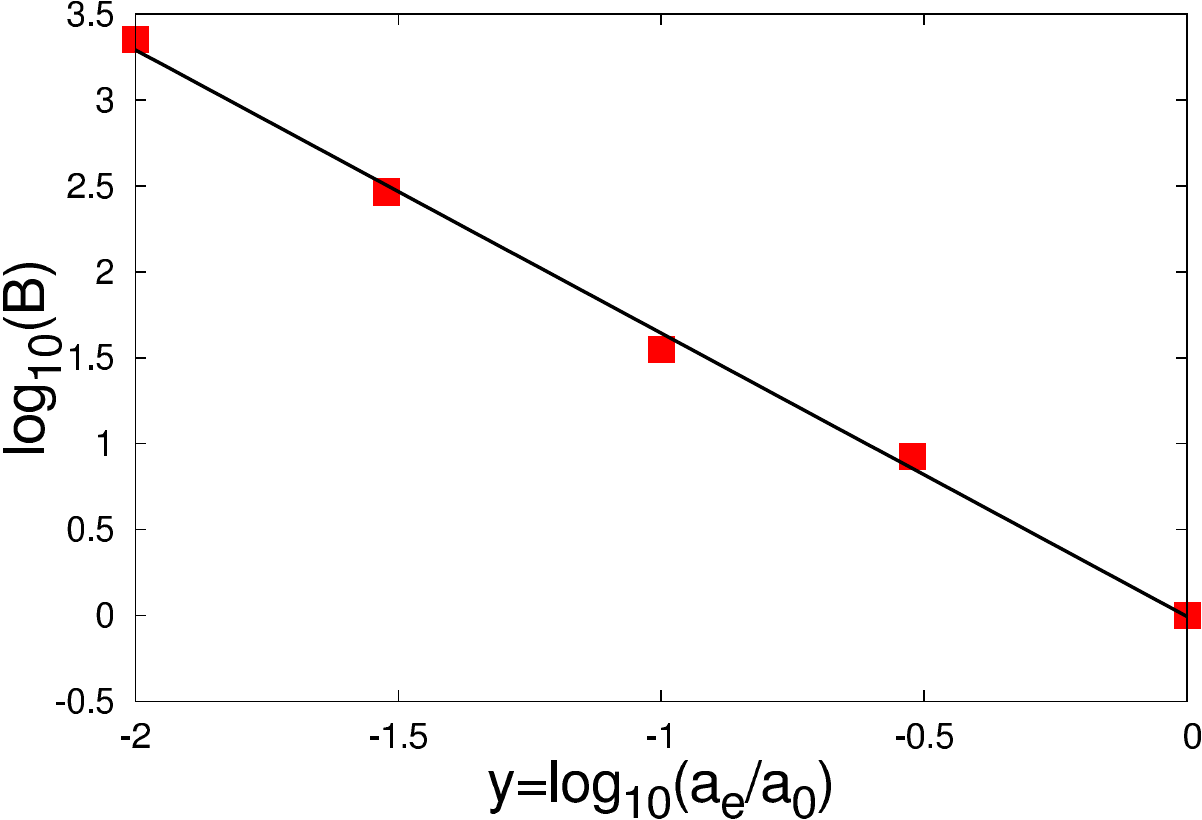}
	\includegraphics[width = 5.6cm] {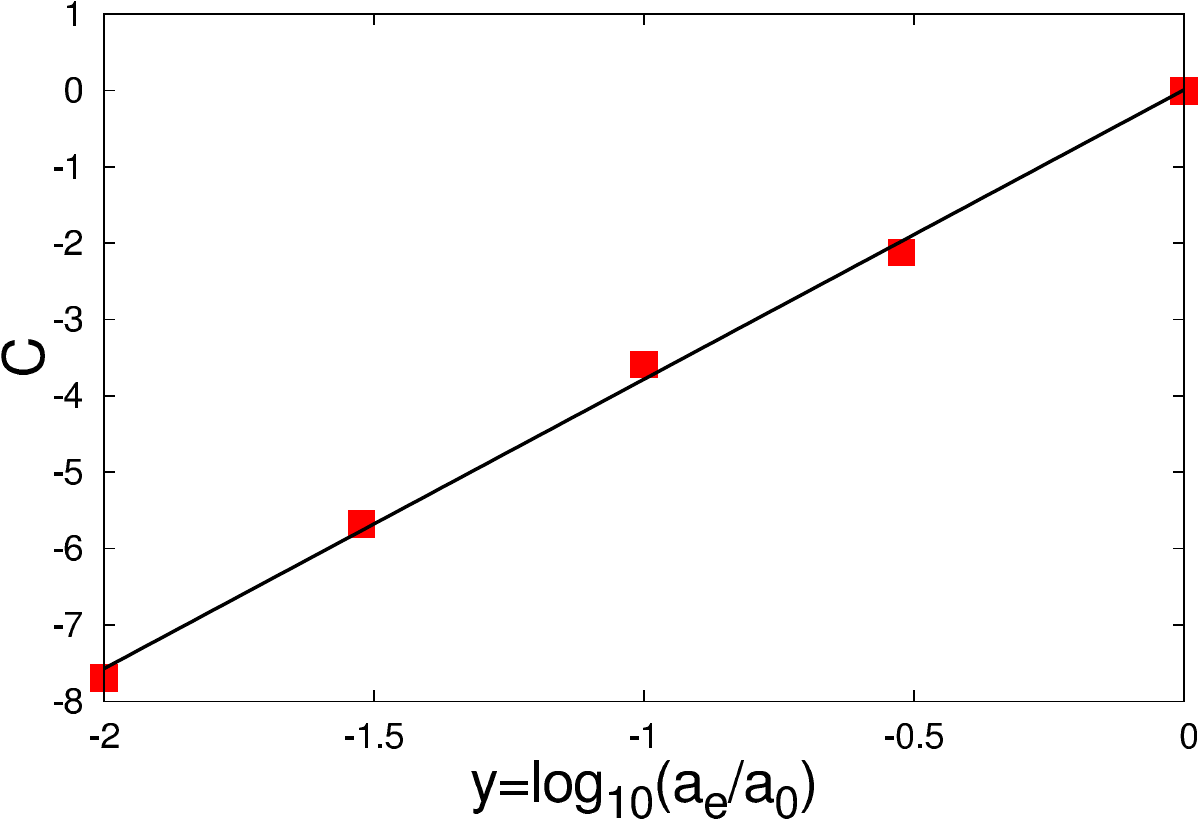}
	\caption{ The best-fitting functions $A, B$ and $C$ to the data in table 1 in \citet{Haghi09} in dependence of the external acceleration ($y$).}
	\label{fit-function}
\end{figure*}

\begin{figure*}
	\centering
	\includegraphics[width = 17.7cm] {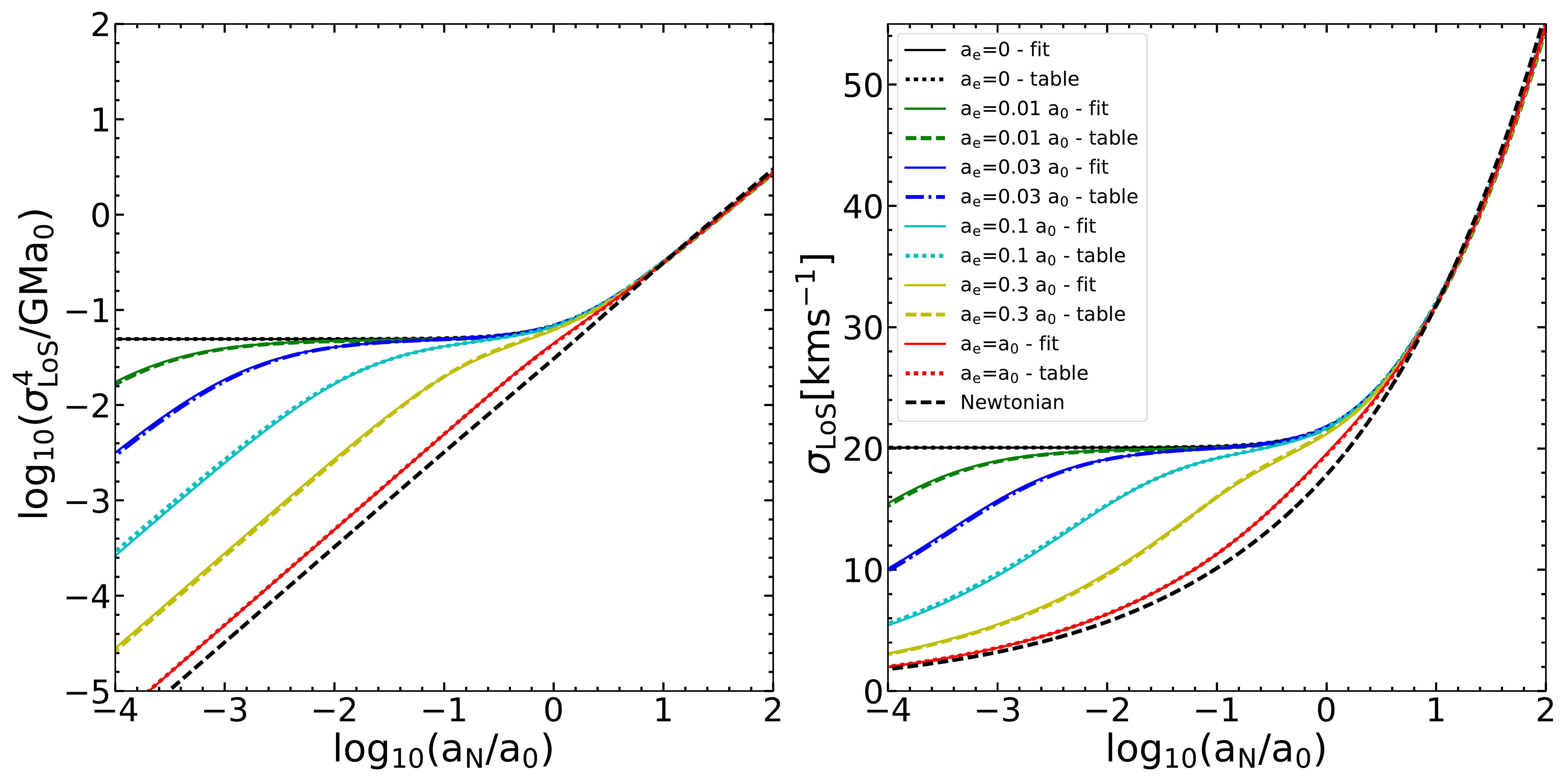}
	\caption{The line of sight velocity dispersion for a stellar system as a function of its internal acceleration when embedded in different external accelerations. The dashed lines show the results using the coefficients given in table 1 of \citet{Haghi09}. The solid lines show the velocity dispersion using our best-fitting functions $A, B$ and $C$ (Equation \ref{Fitting_functions}) to the data in that table. The $x$-axis gives the Newtonian internal acceleration of the system. In order to have different Newtonian accelerations, we vary the half-mass radii but use a fixed mass of $M_{DF2} = 2 \times 10^8 M_\odot$.}
	\label{fit-table}
\end{figure*}

The one-dimensional (line-of-sight) MOND velocity dispersion $\sigma_{\rm M, EF}$ (in km/s) for a system of baryonic mass $M$ and 3D half-mass radius $r_h$ with internal acceleration $a_i \equiv G\,M/\left( 2\,r_h^2\right)$ is
\begin{equation}
{\rm log}_{10} \sigma_{\rm M, EF} = {\rm log}_{10} \sigma_{\rm M} + F(a_e),
\label{eq:EF}
\end{equation}
where the velocity dispersion of an isolated system
\begin{equation}
\sigma_{\rm M} =
\left( {\frac{4}{81}}\, G\, M\, a_0 \right)^{\frac{1}{4}} \times
\left(1 +
0.56\,{\rm exp} \left( 3.02 \, x \right)
\right)^{0.184},
\label{eq:isol}
\end{equation}
$x \equiv {\rm log}_{10}\left( a_i / a_0 \right)$ and
\begin{equation}
F(a_e) = - { \frac{A(a_e)}{4}} \,
\left( {\rm ln}  \left[ {\rm exp}
\left( -{\frac{x}{A(a_e)}} \right) + B(a_e) \right] + C(a_e) \right), \label{eq:fe}
\end{equation}
with $G$ being Newton's gravitational constant. Note that Equations \ref{eq:EF}, \ref{eq:isol} and \ref{eq:fe} supersede the equation used by  \citet{vanDokkum+18c} by a correction factor which ensures the correct behaviour as the internal acceleration rises above $a_0$ and the dynamics become Newtonian. These formulae are chosen because they reproduce well the previous analytical velocity dispersion estimators \citep{Milgrom94, MM13} for systems in the external-field dominated case ($\sigma_{\rm M,EF} = \sqrt{ G\,M\,a_0 / \left( 4\,r_h\,a_e \right)}$) and $\sigma_{\rm M, EF} \approx \sigma_{M} = \sqrt[4]{4\,G\,M \, a_0 / 81}$ for the isolated deep-MOND regime \citep{Milgrom_1995} and the \textsc{N-MODY} results.

We fit the data in \citet[][table 1]{Haghi09} using the functions $A, B$ and $C$ with argument $y \equiv {\rm log}_{10} \left( a_e/a_0 \right)$:
\begin{eqnarray}
A(a_e) &= {\frac{5.3}{\left(10.56 + \left( y +2 \right)^{3.22}\right)}}, \\
B(a_e) &=10^{-(1.65\, y+0.0065)},\\
C(a_e) &= 3.788\,y + 0.006.   \label{Fitting_functions}
\end{eqnarray}

%\begin{align*}
%A(a_e) &= {5.3 \over\left(10.56 + \left( y +2 \right)^{3.22}\right)  }, \\
%B(a_e) &=0.985 \, {\rm exp} \left(-1.65\, y \right),\\
%C(a_e) &= 3.788\,y + 0.006
%%C(a_e) &= -12.83\,
%%{\rm ln}
%%\left(
%%{\rm exp} \left(  {-\left( y + 2 \right)  \over 1.28} \right)
%% + 0.85
%%\right),
%\end{align*}

These fitting functions are shown in Fig. \ref{fit-function}. Therefore, for any non-isolated system with known external acceleration $a_e$, it is possible to calculate  $\sigma_{M,EF}$  using these functions.

To visualize the effect of different external accelerations, we plot the MONDian velocity dispersion as a function of internal acceleration from a weak to a strong external field (Fig. \ref{fit-table}). In order to see how well these fitting functions reproduce the previous results of \citet{Haghi09}, we compare the line-of-sight velocity dispersion found using the best-fitting functions $A, B$ and $C$ (Eq. \ref{Fitting_functions}) with the result of Eq. \ref{eq:EF} using the coefficients given in table 1 of \citet{Haghi09}.

%As can be seen in Figure \ref{fit-table}, the line-of-sight velocity dispersion resulting from our best fitting functions are in excellent agreement with those calculated using the coefficients given in table 1 of \citet{Haghi09}.

%IB altered sentence below slightly.
It should be noted that our three fitting functions are arbitrarily chosen because they reproduce the results of \citet{Haghi09} very well. Although function $A(a_e)$ in Fig. \ref{fit-function} does not perfectly match the simulated data, Figure \ref{fit-table} shows that the line-of-sight velocity dispersion resulting from our best fitting functions are in excellent agreement with those calculated using the coefficients given in table 1 of \citet{Haghi09}. There is at most a 0.1 percent difference between the numerically simulated values in table 1 of \citet{Haghi09}, $\sigma_{table}$ and our analytic fit in Fig. \ref{fit-function}, i.e. $|\sigma_{table} - \sigma_{M,EF}|/\sigma_{M,EF} < 0.001$.

\newpage

\subsection{Comparison with \citet{FM12}}

\begin{figure}
	\centering
		\includegraphics[width =8.5cm] {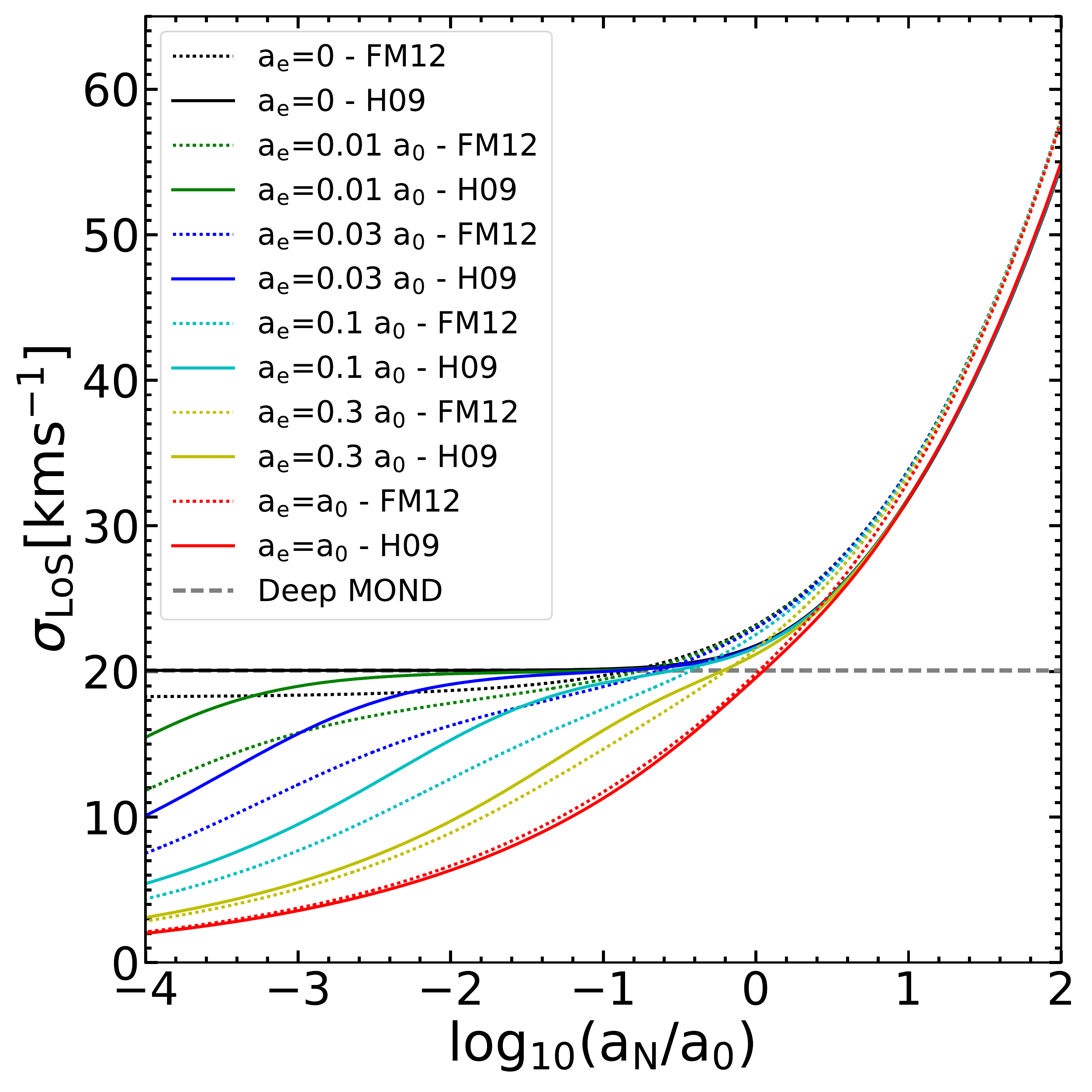}
\caption{ The predicted  global one-dimensional line-of-sight velocity dispersion including the EFE for stellar systems embedded in different external accelerations (different colours) is depicted as a function of internal acceleration using two different approaches. The ansatz proposed by \citet{FM12} is shown by dotted lines and the analytical quantification of the EFE by \citet{Haghi09} using a MOND $N$-body integrator is shown as solid lines. The $x$-axis gives the Newtonian internal acceleration of the system. As expected, the deep MOND velocity dispersion is constant since the mass is fixed in all models.}
	\label{FM12-H09}
\end{figure}

As another check on our analytical formulation of the velocity dispersion, we compare it with the ansatz proposed by \citet{FM12} considering the net MOND effect in one dimension (their eq. 59). When $a_i \approx a_e < a_0$, the object is quasi-Newtonian i.e. Newton's law of Gravity applies with an effective gravitational constant $G_{eff}\geq G$ \citep{FM12, MM13, Famaey18, Muller_2019}. According to the \citet{FM12} formulation,  the MOND acceleration $a$ at the half-light radius is
\begin{equation}
(a + a_e) \, \mu\left(\frac{a+a_e}{a_0}\right) = a_N + a_e \mu \left( \frac{a_e}{a_0} \right) \, ,
\label{efe}
\end{equation}
where $\mu$ is the MOND interpolating function, $a$ is the MONDian internal gravitational field, $a_e$ is the external field from the host, and $a_N=GM/2{r_h}^2$ is the Newtonian internal gravitational field for the mass embedded within the 3D deprojected half-mass radius $r_h$. Using the simple interpolating function \citep{Famaey_2005}, we solve this equation for $a$. The external acceleration can be approximated by $a_e = v^2/D_{sep}$, where $v$ is the rotational velocity of the external galaxy and $D_{sep}$ is the separation between the two objects. Adopting the mass estimator in \citet{Wolf10} to calculate the line-of-sight velocity dispersion as $\sigma_{los}=\sqrt{{G_{eff}M}/\left({6r_h}\right)}$, one can calculate the true velocity dispersion of the system, corrected for the external field. Here, the effective gravitational constant in MOND is defined as $G_{eff} = G_N [a(r_h)/a_N (r_h)]$. In Figure \ref{FM12-H09}, we show the MONDian velocity dispersion as a function of internal acceleration for weak to strong external fields using the formulation of \citet{FM12} (dotted lines) and \citet{Haghi09} (solid lines).

It is important to mention that we calculated $\sigma_{los}$ using the \citet{FM12} relation in the isolated deep-MOND regime for DF2 and obtained a value of ${\approx 18}$ km/s, which is 10\% lower than the deep-MOND prediction of 20 km/s for the isolated system. In the Newtonian regime, the velocity dispersion from \citet{FM12} is ${\approx 5\%}$ higher than the values calculated in our formalism. The \citet{FM12} formula thus leads to a  velocity dispersion smaller by 10\% in the MOND regime and larger by 5\% in the Newtonian regime compared to our analytical formulation. Therefore, the global one-dimensional line-of-sight velocity dispersion of a non-isolated stellar system lying in the intermediate external-field regime probably differs by ${\approx 10-15\%}$ between these formalisms.

This could be due to the different interpolating function used in the MOND $N$-body integrator \citep{Haghi09}. It should be noted that to calculate $\sigma_{\rm M, EF}$ with the \citet{FM12} ansatz, we use the Newtonian mass estimator $\sigma_{los}=0.36\sqrt{GM/R_h}$ in \citet{Haghi09} instead of the \citet{Wolf10} mass estimator $\sigma_{los}=0.41\sqrt{GM/R_h}$). This could be another source of difference between the results of our analytic formulation and the \citet{FM12} ansatz. In any case, the difference can be practically neglected in view of the typical measurement uncertainties.

\section{NGC 1052-DF2}

In this section, the above analytic formulae are applied for the case of DF2 and compared with $N$-body simulations. The {NGC 1052} group has a systemic velocity of $1425\,$km/s with a galaxy--galaxy dispersion of $111$ km/s \citep{vanDokkum+18a}. The main group host galaxy NGC 1052 has a baryonic mass $M_{\rm NGC1052} = 10^{11}\,M_\odot$ \citep{Bellstedt+18} if it lies at a distance of 20 Mpc. If DF2 is at its projected distance to NGC 1052 then the two galaxies are $D_{\rm sep}=80\,$kpc apart, but a more likely distance is $D_{\rm sep}\approx 80\,\sqrt{3/2} = 98$ kpc as the sky plane contains two of the three space dimensions.
%IB_OM: Please check where the 111 km/s figure came from. Word baryonic added.

\citet{vanDokkum+18c} assume that DF2 is located at a distance of $D = 20\,$Mpc in the NGC 1052 group. This implies that the effective radius of its population of 10 GCs is $r_e = 3.1\,$kpc, the stellar body of the galaxy has $r_e \approx 2.2\,$kpc and its absolute V-band magnitude is $M_V = -15.4\,$ mag, corresponding to a luminosity $L_V=1.1 \times 10^8 \, L_{\odot}$. We furthermore assume \citep[like][]{vanDokkum+18c} that the mass-to-light ratio of the stellar population is $M_*/L_V = 2$, while spectroscopy suggests a slightly lower value of 1.6 \citep{Dab16a}. These values for $M_*/L_V$ are in any case well consistent with typical dwarf galaxies in this luminosity range (see e.g. fig. 9 in \citealt{Dab16a}).

\subsection{The inferred velocity dispersion of NGC 1052-DF2}
\label{sec:veldisp}

Using the ten GCs of DF2, \citet{vanDokkum+18b} found $\sigma=7.8^{+5.2}_{-2.2}\,$km/s. Here we revisit the velocity dispersion calculation, which is important for the conclusions as to how much dark matter is contained in DF2 and whether MOND can be falsified using the 10 globular clusters (GCs) with measured radial velocities as was suggested by \citet{vanDokkum+18b}.

We model the true GC radial velocities as following a Gaussian distribution about some mean $\mu$ with intrinsic dispersion $\sigma_{int}$.  This stands in contrast to the biweight distribution \citep{Beers+90} favoured in \citet{vanDokkum+18c} and also used in \citet{Kroupa97}, but a Gaussian/normal distribution is the simpler model and is in fact often realized in nature, since the central limit theorem states that a distribution arising from different random processes approaches the Gaussian distribution. To determine the likelihood of a particular model ($\equiv$ combination of $\mu$ and $\sigma_{int}$), we use the fact that a normal distribution with dispersion $\sigma$ has a probability% (e.g. eq.~58 in \citealt{BZ16})
\begin{equation}
P \propto \frac{1}{\sigma} \rm{e}^{-\frac{\left( Data - Model \right)^2}{2\sigma^2}}.
\label{probability}
\end{equation}

\begin{figure}
	\centering
	\includegraphics[width = 8.4cm] {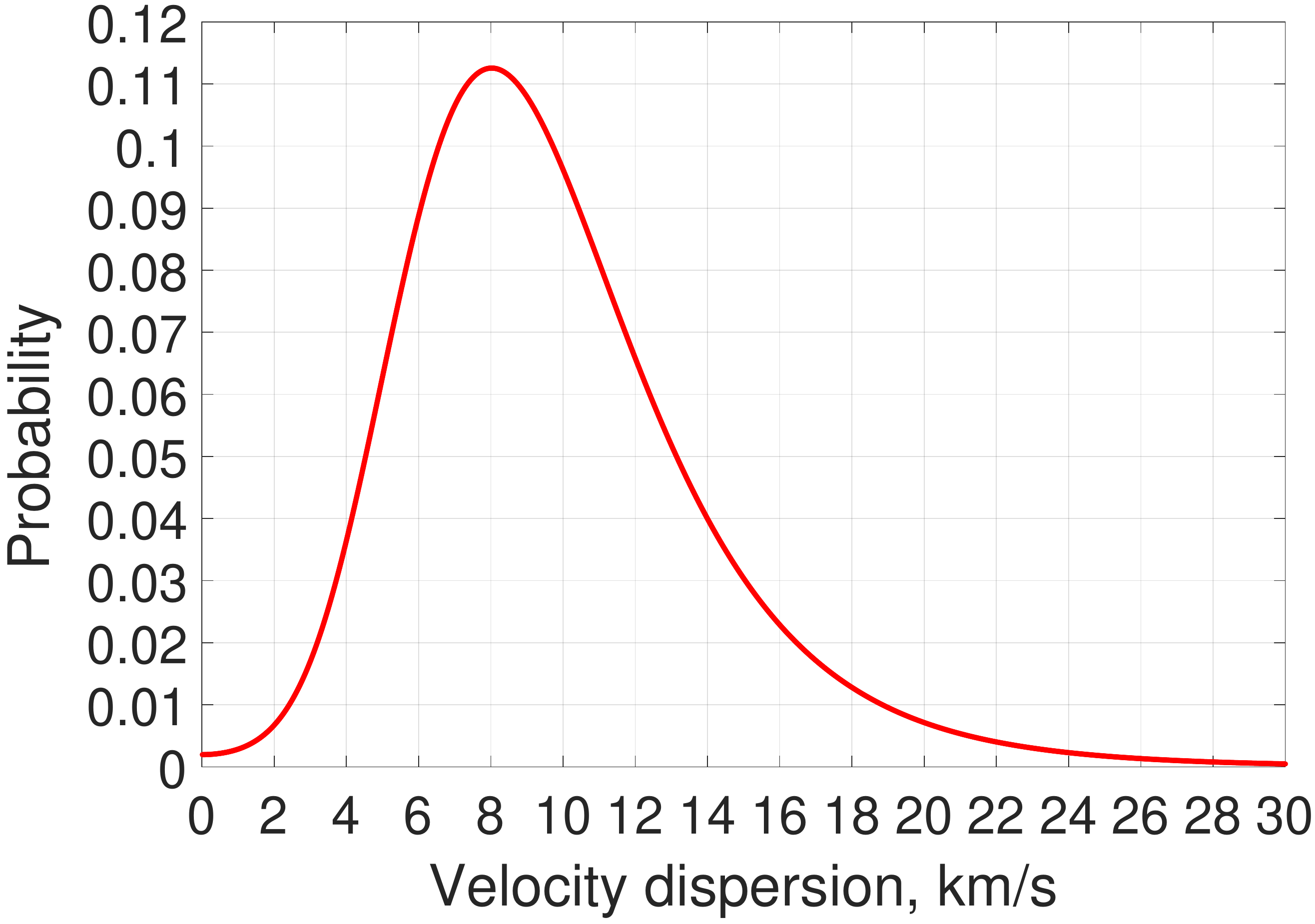}
	\caption{Our marginalized probability distribution for the intrinsic velocity dispersion amongst the ten GCs observed by \citet{vanDokkum+18c}. The most likely dispersion is 8.0 km/s, with a 68.3\% confidence interval of $5.0 - 12.3$ km/s (other confidence intervals are given in Table \ref{Confidence_levels_sigma}).}
	\label{NGC_1052_marginal_sigma}
\end{figure}

A measurement can differ from the mean due to both measurement errors and intrinsic dispersion. Assuming both are Gaussian, we can add these in quadrature. Thus, the probability of a model is
\begin{eqnarray}
P \left( \mu, \sigma_{int} \right) ~&\propto & \prod_{i = 1}^{N} \frac{1}{\sigma_i} \rm{e}^{-\frac{\left( v_{r,obs,i} - \mu \right)^2}{2 {\sigma_i}^2}} \, ,\\
{\sigma_i}^2 ~&=& {\sigma_{obs,i}}^2 + {\sigma_{int}}^2 \, ,
\label{sigma_total}
\end{eqnarray}
where the intrinsic velocity dispersion of the GC system is $\sigma_{int}$ and the uncertainty on the velocity measurement of the i\textsuperscript{\emph{th}} GC is $\sigma_{obs,i}$.   Using this procedure on a grid of values in $\left( \mu, \sigma_{int} \right)$, we obtain the probabilities of different models relative to the most likely model (see Appendix \ref{Appendix} for more details). Marginalizing over the systemic radial velocity $\mu$, we obtain the probability distribution of $\sigma_{int}$ (Fig.~\ref{NGC_1052_marginal_sigma}). At the 68.3\% confidence level, $\sigma_{int} = 8.0^{+4.3}_{-3.0}$ km/s (other confidence intervals are listed in Table \ref{Confidence_levels_sigma}). Due to measurement errors, this is slightly lower than the root mean square dispersion of the radial velocities. However, it is not much lower, as is readily apparent from the data $-$ GCs 39 and 92 have radial velocities differing by~$29\,$km/s but $\sigma_{obs,i} \leq 7\,$km/s in both GCs \citep[][figure 1]{vanDokkum+18c}. This suggests that $\sigma_{int} \approx 8\,$km/s, as confirmed by the present analysis. This revised velocity dispersion shows DF2 to be in excellent agreement with the expected MOND value \citep{Kroupa_2018}. It should be noted that our inferred value agrees quite well with that of \citet{vanDokkum+18b}, which improves on \citet{vanDokkum+18c} by including more data. Moreover, our result is consistent with the most recent estimation of the velocity dispersion of DF2 by \citet{Martin_2018} when allowing for the possibility that some GCs may be interlopers. Similarly to their analysis, we also find no compelling evidence that interlopers affect our inferred $\sigma_{int}$ (Figs. \ref{NGC_1052_contour} and \ref{NGC_1052_subsamples}).

\begin{table}
	\centering
	\begin{tabular}{cc}
		\hline
		Confidence interval & Range in $\sigma$, km/s \\ \hline
		0 sigma (most likely value) & 8.0 \\
		1 sigma (68.3\% confidence) & 5.0 $-$ 12.3 \\
		2 sigma (95.4\% confidence) & 2.4 $-$ 18.8 \\
		3 sigma (99.7\% confidence) & 0.0 $-$ 28.4\\ \hline
	\end{tabular}
	\caption{Our inferred 0, 1, 2 and 3 sigma-equivalent confidence intervals on $\sigma_{int}$.}
	\label{Confidence_levels_sigma}
\end{table}

\subsection{Analytic expectations}

Having inferred the internal velocity dispersion of DF2 (Figure \ref{NGC_1052_marginal_sigma}), its expected dependence on $D_{\rm sep}$ and $M_{\rm NGC1052}$ is shown in Fig.~\ref{fig:MOND} using the analytic formulation of MOND calibrated using numerical simulations (Section \ref{sec:analytic}). For the host mass $M_{\rm NGC1052}=10^{11}\,M_\odot$, the velocity dispersion of DF2 is $\sigma_{\rm M, EF} = 12\,$ and 14 km/s for $D_{\rm sep}=80\,$ and 98 kpc, respectively. This is clearly in agreement with the here inferred velocity dispersion (Sec.~\ref{sec:veldisp}) and the value arrived at by \citet{vanDokkum+18b}, $\sigma=7.8^{+5.2}_{-2.2}\,$km/s, within their~2-sigma confidence range. Another recently determined stellar velocity dispersion of DF2 also agrees with our MOND calculations \citep{Emsellem_2018}. They measured the velocity dispersion of GCs in DF2 by spectroscopic analysis and got a value of $10.6^{+3.9}_{-2.3}$km/s. Moreover, they also studied the velocity dispersion of stars in the dwarf galaxy, obtaining $10.8^{+3.2}_{-4.0}$ km/s. These values agree well with the MOND expectation, as does the recent stellar body measurement of $8.5^{+2.3}_{-3.1}$ km/s by \citet{Danieli_2019}.
%IB deleted instead in the next sentence.

If $D_{\rm sep} > 150\,$kpc, then DF2 would be isolated (unless another major galaxy is in its vicinity) and the expected MOND velocity dispersion would be $\sigma_{\rm M}\approx 20\,$km/s (Eq.~\ref{eq:isol}). This would challenge MOND with just over two-sigma confidence according to the velocity dispersion reported by \citet{vanDokkum+18b,vanDokkum+18c} but is compatible within the three-sigma confidence range of the here obtained velocity dispersion using the same data. DF2 is thus most likely in a quasi-Newtonian state, confirming a central MOND prediction if it is indeed close to NGC 1052 \citep{Kroupa_2018, Famaey18}.

\subsection{\emph{N}-body MOND models}\label{sec:Nbody}

\begin{figure*}
	\includegraphics[width=17cm]{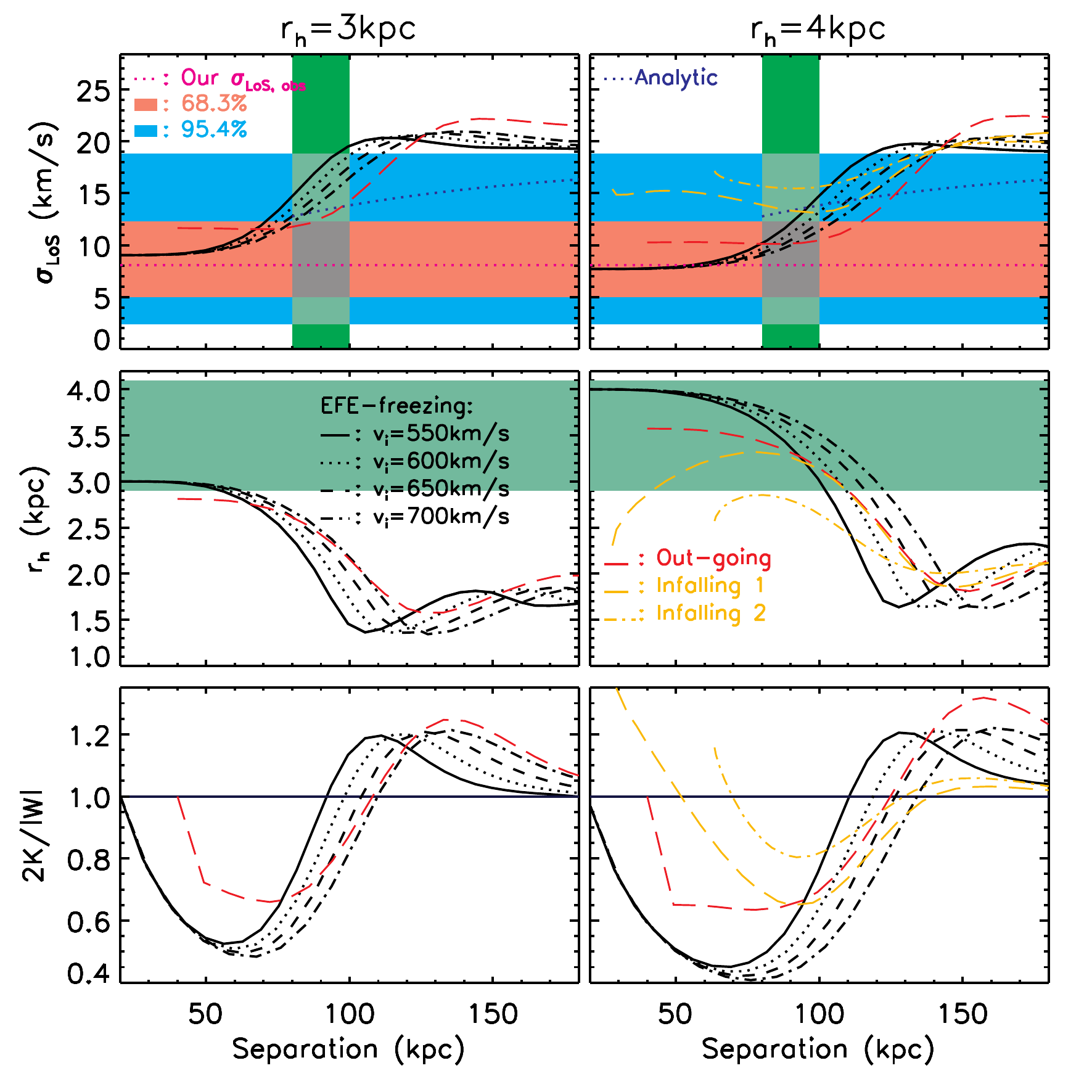} \caption{{\it Upper panels}: The line-of-sight MOND velocity dispersion $\sigma_{\rm M, EF}$ (Eq.~\ref{eq:EF}) is shown as a dotted dark blue line in dependence of the separation $D_{\rm sep}$ between {DF2} and {NGC 1052} for a  baryonic mass of $M_{\rm NGC1052}=10^{11}\,M_\odot$ and $D_{\rm sep} > 80$ kpc. The vertical green region indicates the expanse between the minimum separation $D_{\rm sep}= 80\,$kpc and the most likely distance $D_{\rm sep}= 80\,\sqrt{3/2} \approx 98\,$kpc. The one- and two-sigma ranges on the here constrained velocity dispersion (Section \ref{sec:veldisp}), $\sigma=8.0^{+4.3}_{-3.0}\,$km/s, are shown as horizontal coloured regions. The analytically calculated velocity dispersion $\sigma_{\rm M, EF}$ (only shown for $D_{\rm sep} > 80$ kpc) approaches the isolated value $\sigma_{\rm M} = 20\,$km/s asymptotically for large $D_{\rm sep}$. It declines with decreasing distance and increasing mass because the external field of {NGC 1052} suppresses MONDian self-gravity leading to Newtonian behaviour in the case when $a_i \ll a_e$. For the nominal host mass $M_{\rm NGC1052}=10^{11}\,M_\odot$ \citep{Bellstedt+18} and if $D=98\,$kpc, the MOND velocity dispersion is in agreement within the 2-sigma confidence range of the here measured value. Black lines show the simulated \textsc{N-MODY} line-of-sight velocity dispersion of DF2 as a function of $D_{\rm sep}$ for different orbits with different  pericentre velocities. The left panels represent the simulated dwarf model U1 with an initial half-mass radius $r_h=3\,$kpc and the right panels are for model U2 with $r_h=4\,$kpc. The PoR simulation results are shown as red and gold lines, with model parameters given in Table \ref{PoR_model_parameters}.
	{\it Middle panels}: the sizes $r_h$ of the simulated dwarfs versus $D_{\rm sep}$. The horizontal green region lies between 2.9 kpc and 4.1 kpc, being the deprojected half-light radii of the stars and GCs, respectively. Note how the dwarf contracts when it orbits to a larger $D_{\rm sep}$ because its phantom dark matter halo grows as $D_{\rm sep}$ increases and the external field decreases. This causes the internal acceleration to increase towards the isolated MOND value. {\it Lower panels}: the simulated evolving virial ratio of the dwarf versus $D_{\rm sep}$. Values ${< 1}$ (below horizontal line) imply a deep-freeze state.}
	\label{fig:MOND}
\end{figure*}

Since the size of the ultra-diffuse dwarf galaxy is a few kpc and its velocity dispersion is a few km/s, the crossing time for its stars is $\approx 1$ Gyr. If such a galaxy moves outwards from its pericentre with a high orbital speed, it may not be able to retain dynamical equilibrium when it is far away from the host galaxy, so that it becomes frozen in the quasi-Newtonian regime. In this case, the velocity dispersion is lower than expected if dynamical equilibrium is assumed. This memory effect \citep{Haghi09, WK13a} is considered in the following using $N$-body models.

We model DF2 as being in orbit around a purely baryonic NGC 1052, which we model as an analytical oblate Hernquist profile \citep{Hernquist1991} with axial ratio of 1:1:0.7, a major axis of 2 kpc and  baryonic mass of $M_{\rm NGC1052} = 10^{11}\, M_\odot$ \citep{Bellstedt+18}.

%\beq
%\rho (r) = {M \over 2\pi abc}{1\over r(1+r)^{3}},
%\eeq
%where
%\beq
%r = \sqrt{\left(x\over a\right)^2+\left(y\over b\right)^2+\left(z\over c\right)^2},
%\eeq
%and the constants $a=2.0\kpc,b=2.0\kpc$ and $c=1.4\kpc$ are the typical length scales of the galaxy's major, intermediate and minor axes, respectively.
DF2 is represented using $10^5$ equal-mass particles which are integrated along their orbits using the \textsc{N-MODY} code \citep{NMODY} that considers only a uniform EFE but not tides \citep{WK13a}. The DF2 baryonic mass distribution is a Plummer model, with an overall mass of $M_{\rm DF2}=2\times 10^8\, M_\odot$ with  half-mass radius $r_{\rm h}=3\,$kpc (model U1) and $r_{\rm h}=4\,$kpc (model U2). The equilibrium $N$-body initial conditions are constructed in Newtonian dynamics and then the global velocities of the dwarf particles are increased by the virial ratio $\sqrt{-W/2K}$, where $W$ is the MONDian potential energy and $K$ is the Newtonian kinetic energy \citep{WK13a}.

The pericentre distance is assumed to be $D_{\rm sep} = 20\,$kpc, and this distance along the short axis of the host galaxy is the starting point for the simulations. At this separation, the dwarf has essentially no phantom dark matter halo due to the strong external field close to the host, making its dynamics nearly Newtonian. The initial relative velocity (which is entirely outwards) is varied in the range of $[550,~700]\,$km/s with an interval of $50\,$km/s. The dwarf has a lingering memory of a colder past on an internal crossing time scale $\approx 6 \,{\rm kpc}/8 \, {\rm kms^{-1}} \approx 0.8\,$Gyr.
%IB_OM: may want to clarify that several models were tried, at least for the PoR models - say e.g. that to minimise tidal effects, we tried a larger pericentre. A good match was found when we set this to 40 kpc.

Fig. \ref{fig:MOND} shows that the internal dispersion is essentially frozen at the Newtonian value between $D_{\rm sep}=20\,$kpc and $80\,$kpc. The dwarf would be observed to lack dark matter here. At $D_{\rm sep}=80\,$kpc, the line-of-sight (LoS) velocity dispersion in the simulations is $8.5<\sigma_{\rm M,EF}/{\rm kms^{-1}} < 13.3$ (U1) and  $7.1<\sigma_{\rm M, EF}/{\rm kms^{-1}} < 9.0$ (U2). The virial ratios are below 1 at this separation such that the systems are colder than their equilibrium states, this being the memory effect whereby DF2 is in a deep freeze. The radii of the model dwarfs are 2.1$< r_h/{\rm kpc} <$2.9 (U1) and 3.7$< r_h/{\rm kpc} <$3.9 (U2), both in good agreement with observations of DF2. Its deprojected 3D circularised half-light radius is in the range 2.9$-$4.1 kpc \citep{vanDokkum+18c} if we assume this is 4/3 of the analogous projected quantity \citep{Wolf10}.

As DF2 moves further from NGC 1052, the memory effect disappears near $100< D_{\rm sep}/{\rm kpc} < 130$ (U1) and $120< D_{\rm sep}/{\rm kpc} < 150$ (U2). These simulations show that an orbit with a higher velocity can freeze DF2 out to a larger separation. The deep-freeze state can be recognized when $2K/\left| W \right|<1$. Moreover, an initially more diffuse dwarf can be frozen in the quasi-Newtonian regime out to a larger separation. This demonstrates a competition between the orbital time and the crossing time, and an initial ultra-diffuse galaxy moving rapidly from the inner region near a host galaxy can be frozen in quasi-Newtonian dynamics even at a large separation. The observed DF2 may thus be the first example of the memory effect in MOND.

We have also realised two fully self-consistent simulations (including tides and the EFE) with the adaptive-mesh refinement MOND code Phantom of RAMSES \citep[PoR,][]{Lueghausen14b} of a U1 and a U2 model. These are launched on a hyperbolic orbit from a larger pericentre at $D_{\rm sep}=40\,$kpc to avoid strong tidal effects from the host galaxy. The initial relative velocity is 550 km/s perpendicular to the line of sight, implying an apocentre of 1540 kpc. All our galaxy models consist of live particles. To simplify the calculations, the host galaxy, NGC 1052, has a spherical Plummer density profile with a mass of $10^{11} M_{\odot} $ and a half-mass radius of $1.3$ kpc. Both the tidal and external field are taken into account in these simulations. The model dwarfs are initially in equilibrium at their pericentres (i.e. $D_{\rm sep}=40$ kpc) with cut-off radii of 10 kpc. After launch, our model dwarf galaxies are temporarily frozen in the quasi-Newtonian state (bottom panel of Figure \ref{fig:MOND}). At a distance of ${80-100}$ kpc, $12.1 < \sigma_{\rm M, EF}/{\rm kms^{-1}} < 14.1$ for the U1 model and $12.8 < \sigma_{\rm M, EF}/{\rm kms^{-1}} < 14.7$ for the U2 model, consistent with our \textsc{N-MODY} simulations which only include the EFE.  The parameters of the PoR models are listed in Table \ref{PoR_model_parameters}.

%At such a distance the dwarf galaxy is frozen in MOND regime when it is near the pericenter but in Newtonian regime when it is moving outwards.

For completeness of our analysis, we calculate two PoR models launched from a larger distance of 200 kpc with initial relative velocity of 400 km/s directed such that the pericenter is at 28 (64) kpc. These are shown by the gold dashed (dot-dashed) curves in Fig. \ref{fig:MOND}. We use a starting point 200 kpc away because this makes DF2 almost isolated initially. The external field from the host galaxy is $\approx 0.05 a_0$ such that a more distant starting point would not make any difference to the external field and tidal effects. In these models, the initial half-mass radius of DF2 is 1.5 (2.0) kpc. The size of the system expands when the dwarf galaxy is near pericenter, compensating for the smaller size of the initial model and matching the observed radius. An even better match could be obtained for a larger initial size, which would somewhat lower the velocity dispersion and make this more in line with observations.

%The orbits are launched from outer region of NGC1052. The initial distance is 200kpc and the initial relative velocity is -400km/s.

The infalling satellite puffs up (compared to the outgoing cases) and shows a significant increase in the virial ratio, possibly not surviving a second passage. The system is frozen in the MOND regime near pericenter i.e. it has a virial ratio $>1$. At a separation of 80-100 kpc, the velocity dispersion agrees with the observations within their 2-sigma error range. When the separation is beyond 100 kpc, the prediction from the new model is very similar to those of EFE-only (\textsc{N-MODY}) models.

The GCs are more spread out than the stars of the stellar body of DF2, so the GC velocity dispersion could be slightly lower than our analytical and numerical predictions. This issue does not arise for the stellar velocity dispersion measurements, which are already weighted by luminosity and thus nearly mass-weighted, as in our calculations.

While these experiments are still idealized, they demonstrate that a systematic study of such dwarf satellite galaxies is needed before drawing conclusions about fundamental theory. A more detailed paper is in preparation (Wu et al.) in which multiple orbits for DF2 will be studied.

In summary, the analytically calculated value of $\sigma_{\rm M, EF}$ is verified by \textsc{N-MODY} (a spherical particle-mesh code) and PoR (an adaptive-mesh refinement code) simulations. All our theoretical estimates are consistent with the measured velocity dispersion of DF2.
%IB slightly altered previous sentence.

\begin{table*}
	\centering
	\begin{tabular}{cccccccc}
		\hline
		Model        &   Line type      & Pericenter&Starting position& Starting separation   & Initial relative speed & Initial $r_h$      & Speed at pericenter \\ \hline
		Out-going 1  & Red  dashed     &  40 kpc   &(0, 0, 40) kpc   & 40 kpc                &  550 km/s                 & 3.0 kpc &550 km/s \\
		Out-going 2  & Red  dashed     &  40 kpc   &(0, 0, 40) kpc   & 40 kpc                &  550 km/s                 & 4.0 kpc & 550 km/s \\
		Infalling 1  & Gold dashed     &  28 kpc   &(40, 0, 195) kpc & 200 kpc    &  400 km/s                 & 1.5 kpc            & 553 km/s\\
		Infalling 2  & Gold dot-dashed &  64 kpc   &(80, 0, 183) kpc & 200 kpc    &  400 km/s                 & 2.0 kpc            & 487 km/s \\ \hline
	\end{tabular}
	\caption{The initial parameters of our four PoR models in Fig. \ref{fig:MOND}. The viewing direction is the $x$-axis. The `out-going' models are started at pericenter.}
	\label{PoR_model_parameters}
\end{table*}
%IB changed table label.

\subsection{The properties of NGC 1052-DF2 at different distances} \label{sec:Distance}

The above discussion assumed that DF2 is a physical member of the NGC 1052 group with a projected separation of $\approx 80$ kpc from NGC 1052 which is assumed to be 20 Mpc away (see Sec. 3.4.4). The arguments in favour of this are (i) the line-of-sight velocity is $+378\,$km/s ($3.4\,\sigma$) with respect to the NGC 1052 group and $+293\,$km/s with respect to NGC 1052 and (ii) the non-detection of a gas component in DF2 \citep{Sardone_2019, Chowdhury2018}, suggesting it is part of a galaxy group \citep{Geha06}. In addition, the Hubble Space Telescope should have been able to resolve the red-giant-branch stars if this dwarf galaxy is closer than 10 Mpc, unless its stellar population is non-canonical.
%IB_OM: deleted , (ii) the surface-brightness fluctuation and Planetary Nebula Luminosity Function distance estimates of 20 Mpc \citep[][respectively]{vanDokkum+18c, Fensch_2018}

The surface-brightness fluctuation method can yield inaccurate results because it relies on the number of giant stars per unit surface area \citep{JR01}. This depends on the age and metallicity of the stellar population, the mass distribution of which also depends on the metallicity and star-formation rate \citep{Kroupa13, Yan+17, Jerabkova18}. Galaxies with a high star formation rate are known to be producing stellar populations overabundant in massive stars \citep{Gun11}, while galaxies with a low star formation rate show a deficit of massive stars \citep{Lee09, Watts2018}. Old, dormant galaxies also show significant variations of their stellar populations: elliptical galaxies may be dominated by very low mass stars \citep{vanDokkum10}, while faint diffuse dwarf galaxies have a deficit of low mass stars \citep{Gennaro+18} when compared to the canonical stellar population \citep{Yan+17, Jerabkova18}.

In the following we discuss the properties of DF2 if it were at a shorter distance from Earth, by considering its dynamical $M/L_V$  ratio and the properties of its GC system. Notably, we seek to illuminate how hard the evidence for a $D=20$ Mpc distance is and how much leeway we have for this dwarf to be at about 10 to 13 Mpc. A particularly important question we seek to touch is which major galaxy (NGC 1052 or NGC 1042) along the line of sight can be the host galaxy to which DF2 is a possible satellite.  Can NGC 1052 itself be much closer, such that the NGC 1052, DF2 and possibly DF4 (Sec. 4) system of galaxies might be gravitational bound and at a distance of 10-13 Mpc?  The major tension with this suggestion would be the high peculiar velocity such a distance would imply, and so we discuss other precedence cases of correlated galaxy populations which have group radial velocities which deviate from the Hubble flow significantly (Sec. \ref{sec:systemic}).

\subsubsection{Globular cluster population and specific globular cluster frequency}

That the distance of {DF2} may differ significantly from $20\,$Mpc is indicated by its ten GCs all being ${\approx 4 \times}$ brighter and ${\approx 2 \times}$ larger than the GCs of other known galaxies \citep{vanDokkum+18b}. The GC luminosity function of all known galaxies universally peaks at $M_{\rm V}=-7.7\,$ \citep{Rejkuba12}, while that of DF2 peaks at $M_{\rm V} = -9.1$ for a distance of $D=20$ Mpc. If DF2 were to lie at $D=8$ Mpc, its GCs would appear normally bright and would have radii consistent with normal GCs.

The number of GCs per luminosity of the host galaxy, the specific frequency, is known to increase with decreasing luminosity of the dwarf galaxy for early-type (i.e. dormant) spheroidal galaxies. The specific GC frequency, $S_N = N_{\rm GC}\, 10^{0.4 (M_{\rm V} + 15)}$, is a measure of the number of GCs possessed by a galaxy with absolute V-band magnitude $M_{\rm V}$ \citep{Elmegreen_1999, Georgiev_2010, WK13}.

At $D=20\,$Mpc, with $N_{\rm GC}=10$ GCs and  absolute V-band magnitude of $M_{\rm V}=-15.4$ \citep{vanDokkum+18a}, {DF2} has $S_{\rm N}=6.9$, which is normal for an early-type dwarf galaxy \citep[][figure 3]{Georgiev_2010}. For the distance range in which {DF2} would have normally-bright GCs ($8 < D/{\rm Mpc} < 13$, Fig.~\ref{fig:ML}), the absolute V-band magnitude is between $M_{\rm V} = -13.3$ and $-14.3$ such that $S_{\rm N}$ is between 48 and 19, respectively. The $S_{\rm N}$ value is thus compatible with normal late-type dwarf galaxies at both distances (20 and 10 Mpc; fig. 3 in \citealt{Georgiev_2010}).

\subsubsection{The systemic line-of-sight velocity of {NGC 1052-DF2}} \label{sec:systemic}

%IB_OM: reference added. Candlish_2016 Kogut93
The systemic line-of-sight velocity of its GCs is $1803$ km/s \citep{vanDokkum+18c} such that if they are bound to {DF2} and this velocity were due to the Hubble flow, then $D\approx 20\,$Mpc. The {NGC 1052}-group systemic velocity is $\approx 1400\,$km/s, suggesting physical association and a similar distance. Can {DF2} nevertheless be a foreground dwarf galaxy with for example $D\approx 13$ Mpc? \citet{Trujillo2018} carried out an analysis of all extant stellar-population related  data and showed that these do suggest a much shorter distance (13 Mpc) than previously indicated (20 Mpc). With this revised distance, the galaxy appears to be a rather ordinary low surface brightness galaxy because the luminosity and structural properties of its GCs are similar to those of other galaxies (Section \ref{sec:MLrat}).

The peculiar velocity of DF2 relative to the CMB reference frame would be rather large at
$D\approx 13$ Mpc \citep[$v_{pec}=640 \pm 25$ km/s,][]{Trujillo2018}. Observationally, such a peculiar velocity is not out of the question in the real Universe as the Local Group of galaxies has $v_{pec} = 630$ km/s \citep{Kogut93}. Another high-velocity system is the Leo-I group of galaxies, which lies at a distance of $\approx 10$ Mpc and has a line-of-sight group velocity of $\approx 1000$ km/s, about 300 km/s ahead of the Hubble velocity $v_{_H}$ \citep{Muller18}. Other examples can be found in the vicinity of the Local Group, where galaxies are receding significantly faster than the Hubble expansion (figure 5 in \citealt{McConnachie12}; \citealt{PM14}; \citealt{BZ18}).

%IB: paragraph above and below rephrased slightly.

%For the case at hand, a distance of 13 Mpc implies a peculiar velocity of $\approx 900\,$km/s ($=1800-v_{\rm H}$). Indeed, the radial (and thus peculiar) velocity of {DF2} would be $220\,$km/s less if measured with respect to the cosmic microwave background rather than the Sun.

Systems with $v_{pec} \ga 600$ km/s are unlikely in a $\Lambda$CDM cosmology but very natural in a MOND cosmology \citep[figure 14 of][]{Candlish_2016}. The peculiar velocity of DF2 could have been enhanced if it was flung out by a three-body interaction, perhaps between it, another putative major galaxy at $D\approx 13\,$Mpc (e.g. NGC 1042) and a smaller galaxy. If {DF2} was flung out away from us, its higher radial velocity than {NGC 1052} does not necessarily imply a larger distance. For example, moving at $300\,$km/s for $500\,$Myr implies motion by only $150\,$kpc. {NGC 1042} does have a disturbed morphology, suggesting that it was involved in a relatively recent event \citep{vD19}. It is therefore possible that {DF2} is an isolated normal dwarf galaxy which by coincidence lies along the line of sight to the {NGC 1052} group with a comparable line-of-sight velocity, making it appear extraordinary.

According to Fig.~\ref{fig:ML}, {DF2} becomes a normal dwarf galaxy with baryonic mass $\approx 5 \times 10^7\,M_\odot$ if $8 < D/{\rm Mpc} < 13$ \citep{Mateo_1998, Martin_2008, McConnachie12}. In this case, it may be an isolated dwarf galaxy whose MOND velocity dispersion would be $\sigma_{\rm M}\approx 14\,$km/s, within the 2$\sigma$ confidence range of the observed velocity dispersion of its GCs (Table \ref{Confidence_levels_sigma}). The high line-of-sight systemic velocity of {DF2} could be a chance superposition with {NGC 1052} if NGC 1052 is at a distance of 20 Mpc(see Sec. 3.4.4). In this case, a plausible scenario is that DF2 is a normal satellite of NGC 1042, which may be closer to Earth than NGC 1052. At a distance of 13 Mpc, the sky-projected separation of NGC 1042 and DF2 would be only 78 kpc \citep[][figure 4]{vD19_DF4}.  However, the EFE of this galaxy is insignificant for the internal dynamics of DF2 since the baryonic mass of NGC 1042 is about one order of magnitude lower than that of NGC 1052.

\subsubsection{Implications for the dark matter content of DF2  of a revised distance scale} \label{sec:MLrat}

A fixed $M/L_V$ implies $D$ has no effect on the Newtonian gravity $g_{_N}$ at the effective radius $r_e$. This is because a fixed angular size implies $r_e \propto D$ while a fixed apparent magnitude implies $M \propto D^2$ at fixed $M/L_V$, thereby causing a cancellation between the changes to $M$ and those to $r_e$ under an inverse square gravity law. As equilibrium requires $\sigma^2 \propto r_e \times g(r_e)$ where $g$ is the true gravity, any theory uniquely linking $g$ and $g_{_N}$ has the property that $\sigma \propto \sqrt{D}$. This is true even in the presence of an external field because this is independent of $D$, if we assume that $D_{\rm sep}/D$ remains constant and the mass of the external field's source also scales as $D^2$, which is valid for a fixed $M/L$.
%IB_OM: second sentence added. M_NGC1052 deleted as this is a general statement, not unique to NGC 1052.

\begin{figure}
	\includegraphics[width=8.4cm]{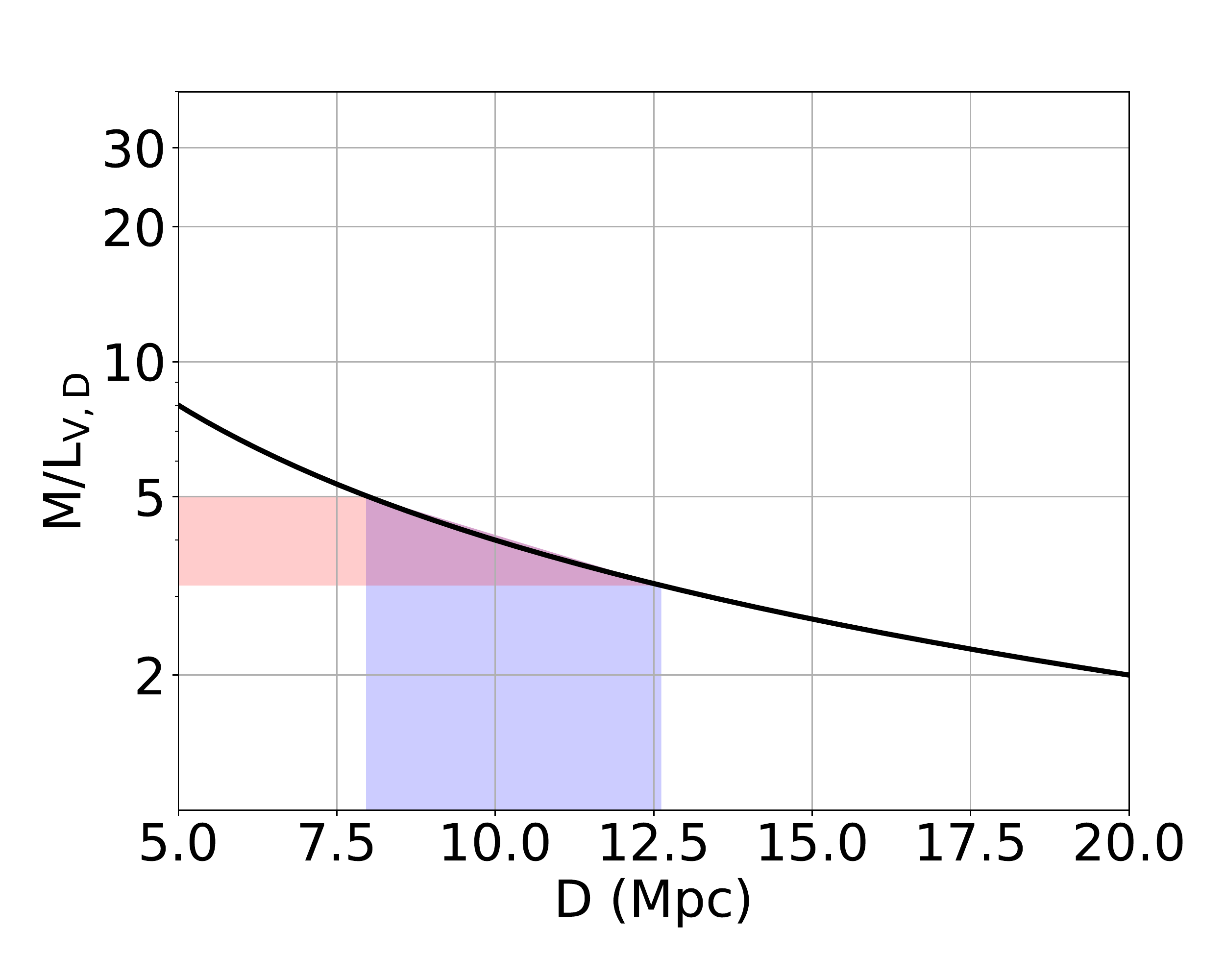}
	\caption{The dependence of the dark-matter content on the distance. The V-band dynamical mass-to-light ratio $M/L_{\rm V, D}$ of {DF2} is shown in dependence of its distance $D$ (Eq.~\ref{eq:MLrat}). If the $N_{\rm GC} = 10$ GCs are comparable in luminosity to those of other galaxies, then $8 < D / {\rm Mpc} < 13$ (indicated by the coloured region), $0.8 < r_e/{\rm kpc} < 1.4$, $1.8 \times 10^7 < L_{\rm V, D} / L_{\odot,V} < 4.6\times 10^7$ and $3 < M/L_{\rm V, D} / (M_\odot/L_{\odot, V}) < 5$, making {DF2} a dwarf galaxy comparable to the bright Local Group dwarf spheroidal satellite galaxies such as Fornax \citep{Cole+12}. The current 3-sigma upper limit on the velocity dispersion is $19.7$ km/s \citep{vanDokkum+18c}, corresponding to an isolated baryonic MOND mass of $M=1.8 \times 10^8\,M_\odot$ (Eq.~\ref{eq:isol}). For a stellar $M/L_{\rm V}=2$, this implies $L_{\rm V}=9 \times 10^7\,L_{\rm \odot, V}$. If {DF2} would have $D > 18$ Mpc and be isolated, then it would constitute a significant MOND outlier due to the absence of the EFE, in addition to its size and GCs making it a very unusual ultra-diffuse galaxy.}
	\label{fig:ML}
\end{figure}

\citet{vanDokkum+18c} calculate the gravitating mass of {DF2} to be $M_{\rm DF2} \approx 2 \times 10^8\,M_\odot$ assuming $\sigma = 3.2\,$km/s. From \citet{Wolf10}, $M_{\rm DF2} \propto \sigma^2 \, r_{\rm e}$ where $\sigma = 7.8\,$km/s is the line-of-sight velocity dispersion of the GCs in  {DF2} measured by \citet{vanDokkum+18b}, $r_e= \theta \, D$ is the effective radius of the GC system and $\theta = 31.84\arcsec =1.54 \times 10^{-4}$ is the angular radius on the sky. If {DF2} lies at a distance $D$, then its absolute V-band luminosity becomes $L_{\rm V, D} = \left( D/20\,{\rm Mpc} \right)^2 \; L_{\rm V,   20\,Mpc}$, where $L_{\rm V, 20\,Mpc} = 1.1 \times 10^8 \,L_{\odot, V}$ at a distance of 20 Mpc \citep{vanDokkum+18c}. As they obtained $M/L_V = 2$, the mass-to-light ratio in Solar units at distance $D$ is
\begin{eqnarray}
{\frac{M}{L_{\rm V,D}}} ~=~ 2 \left(D / {\rm 20\, Mpc} \right)^{-1}.
\label{eq:MLrat}
\end{eqnarray}

A smaller $D$ would imply a smaller luminosity and effective radius. This would increase the V-band dynamical mass-to-light ratio $M/L_{\rm V, D}$ of its GC system for their observed velocity dispersion. Assuming this is 8 km/s, Fig.~\ref{fig:ML} shows how the Newtonian dynamical $M/L_V$ ratio changes with distance.

% The Newtonian dynamical $M/L_V$ shown here is based on an internal velocity dispersion of 8 km/s. %Slightly changed by IB. 8.0 is too precise here.

%being ${M \over L_{\rm V,D}}=2\, {M_\odot \over L_{\rm \odot, V}}$ as used by van Dokkum et al. \citep{vanDokkum+18}.

%In MOND, a fixed $M/L_{\rm V}$ implies $M_{\rm DF2} \propto D^2$ such that $\sigma_{\rm M} \propto \sqrt{D}$.

\subsubsection{Implications for MOND of a revised distance scale}
\label{Distance_implications_DF2}

%\citep{Trujillo2018} recently claimed that they detected the tip of the DF2 red giant branch using Hubble Space Telescope observations. Those authors use this feature to estimate a distance of ${13.0 \pm 0.4}$ Mpc, confirming the many lines of evidence discussed in this section. We therefore address the implications for MOND of a reduced distance.

In general, scaling the distance to all relevant objects by some factor $a$ affects the velocity dispersion by $\sqrt{a}$ because the external and internal gravitational fields remain constant if the $M/L$ values are held fixed (Section \ref{sec:MLrat}). If NGC 1052 is assumed to be at 20 Mpc while DF2 is at 13 Mpc, then DF2 becomes an isolated object. Using $\sqrt{13/20} \approx 0.8$, we see that the MOND prediction becomes $\sigma_{M} = 16$ km/s. Within the 2$\sigma$ confidence range, this is consistent with the here inferred dispersion (Table \ref{Confidence_levels_sigma}), the $\sigma=7.8^{+5.2}_{-2.2}$ km/s measurement of \citet{vanDokkum+18b} and the DF2's GC velocity dispersion of $10.6^{+3.9}_{-2.3}$ km/s obtained by \citet{Emsellem_2018} based on eleven GCs.  The latter workers also studied the velocity dispersion of stars in DF2, inferring a dispersion of $10.8^{+3.2}_{-4.0}$ km/s. All these values agree with the MOND expectation for an isolated DF2 at 13 Mpc, though the $8.5^{+2.3}_{-3.1}$ km/s measurement by \citet{Danieli_2019} is uncomfortably low for MOND.

%IB_OM: Reference to previous section added.

However, DF2 need not be isolated if it is  13 Mpc from Earth. The gas-poor nature of DF2 \citep{Chowdhury2018, Sardone_2019} suggests that it may be in or was in a galaxy group.  A possible candidate host galaxy is NGC 1052, whose distance is far from certain. \citet{Theureau_2007} reported a value of ${17 \pm 3}$ Mpc based on the Tully-Fisher relation \citep{Tully_Fisher_1977}. In a MOND context, this relation is a consequence of fundamental physics for isolated systems \citep{Milgrom83b, McGaugh2000}.
%It is thus expected to work rather well for a high surface brightness galaxy like NGC 1052 since it should be fairly immune to the EFE, though the uncertain inclination of NGC 1052 makes it more difficult to be certain of the results.
Thus, it is quite possible that DF2 and NGC 1052 have a similar distance of $D{\approx 13}$ Mpc.
%IB_OM: added mention of inclination uncertainties after the comma.

If we assume that the distance to both NGC 1052 and DF2 is reduced by the same factor, then the numerical results in Section \ref{sec:Nbody} can easily be scaled to a lower distance. The main consequence is that the calculated $\sigma_{M,EF}$ is reduced by a factor of 0.8. The timescales of the simulations and all velocities in them would also be reduced by the same factor, while the mutual separation  would be reduced by $0.8^2=0.64$ and the masses by $0.8^4=0.41$. A 20\% reduction in the predicted $\sigma_{M,EF}$ makes our results even more consistent with observations, especially if these eventually favour the lower value reported by \citet{Danieli_2019}.
%IB_OM: Please check Famaey's latest MOND predictions for DF2 and if it can work at 13 Mpc if the host galaxy is NGC 1042. We could then say that having verified our analytic formula with N-body simulations, previously a similar formula was used to show that the lower mass host NGC 1042 would create a sufficient EFE if combined with the assumption that both are only 12 Mpc away.

\section{NGC 1052-DF4} \label{sec:DF4}

\citet{vD19_DF4} announced the discovery of DF4, a second galaxy lacking dark matter with rather similar properties to DF2 in terms of its size, surface brightness, morphology and distance. They measured the root mean square spread of observed radial velocities as $\sigma_{obs}=5.8$ km/s amongst a population of seven luminous globular clusters that extend out to a distance of 7 kpc from the centre of DF4. Taking observational uncertainties into account, they determined an intrinsic velocity dispersion of $\sigma_{intr}=4.2^{+4.4}_{-2.2}$ km/s. Using  our analytic formulae for the global one-dimensional line-of-sight velocity dispersion $\sigma_{M,EF}$ of a non-isolated stellar system lying in the intermediate external field regime (Section \ref{sec:analytic}), we calculate the expected internal velocity dispersion of DF4 in MOND.

\subsection{The external field effect of three possible hosts}

Since the group environment of DF4 may host several large galaxies, we have to consider the EFE of all influencing galaxies when calculating the MONDian velocity dispersion of DF4. There are three other bright nearby galaxies in the group close to DF4 in terms of their sky positions. These galaxies are NGC 1052, NGC 1042, and NGC 1035 with baryonic masses of $M_{\rm NGC1052}=10^{11}\,M_\odot$ \citep{Bellstedt+18}, $M_{\rm NGC1042}=2.4\times 10^{10}\,M_\odot$, and $M_{\rm NGC1035}=1.9\times10^{10}\,M_\odot$ \citep{Muller_2019}, respectively, assuming they are all at a distance of $D=20$ Mpc. To estimate the total baryonic mass of NGC 1042 and NGC 1035 with K-band absolute magnitudes of $M_{K, 1042}=-8.85$ and $M_{K, 1035}=-9.13$, respectively \citep{Skrutskie_2006, Brough_2006},  we follow the approach in \citet{Muller_2019}. We first convert the K-band luminosities to stellar masses using a $M/L_K$ ratio of 0.8. Then we add the gas mass using eq. 2 from \citet{DiCintio_2016}.

\begin{figure}
	\includegraphics[width=8.5cm]{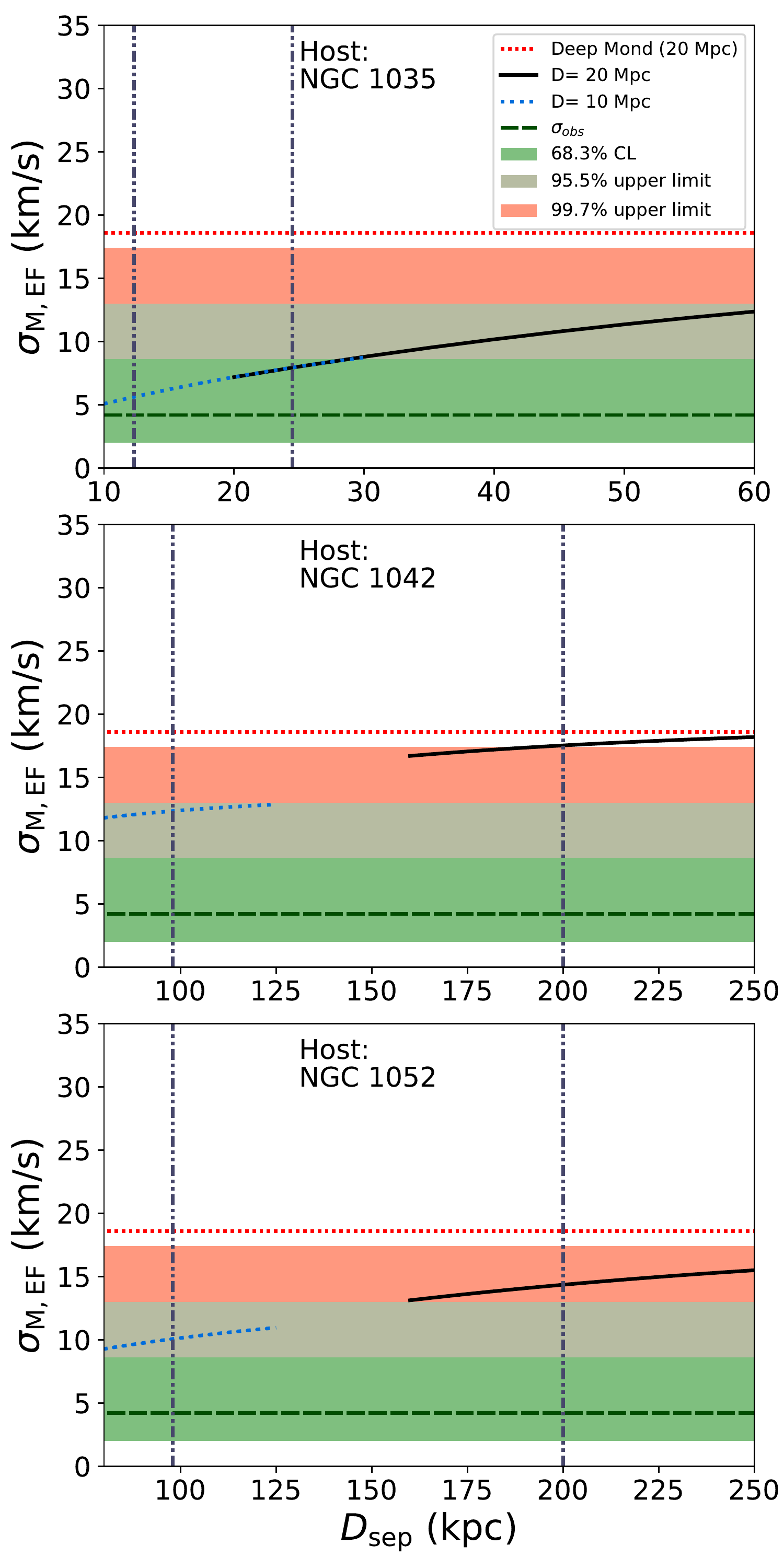}
	\caption{The line-of-sight MOND velocity dispersion of DF4 ($\sigma_{\rm M, EF}$, (Eq.~\ref{eq:EF}) in dependence of the separation $D_{\rm sep}$ between it and {NGC 1035} (top), {NGC 1052} (middle) and {NGC 1042} (bottom) for two different distances of $D=20$ and 10 Mpc (solid and dotted lines, respectively). The vertical dashed-dotted lines indicate the most likely separation  $D_{sep}= D_{proj}\sqrt{3/2}$ for each assumed $D$. The 1, 2 and 3 sigma ranges on the measured velocity dispersion \citep[$\sigma=4.2^{+4.4}_{-2.2}$ km/s,][]{vD19_DF4} are shown as horizontal coloured regions. The analytically predicted velocity dispersion $\sigma_{\rm M, EF}$ (only shown for $D_{\rm sep} \geq D_{\rm proj}$) approaches the isolated value ($\sigma_{\rm M} = 18.6$ km/s) asymptotically for large $D_{\rm sep}$. It declines with decreasing distance due to the EFE.}
	\label{DF4}
\end{figure}

We therefore evaluate the MONDian velocity dispersion of DF4 by considering the EFE on it associated to each bright galaxy separately. In reality, the EFE from all three objects should be considered, but this is left to future work (though we briefly touch on this in Section \ref{Combined_EFE}). Here, we only consider the separation $D_{sep}$ between DF4 and its possible host, allowing implicitly the distance, $D$, of DF4 from Earth to be significantly uncertain. The expected dependencies of the internal velocity dispersion of DF4 on $D_{sep}$ between it and {NGC 1052}, {NGC 1042} and {NGC 1035} are shown in Figure \ref{DF4}.

We show the effect of a reduced distance $D$ (dotted blue lines) based on halving the distance to all relevant objects. This simply involves redrawing the curves with $D_{sep} \to D_{sep}/2$ and $\sigma \to \sigma/\sqrt{2}$ (Section \ref{Distance_implications_DF2}). For each assumed distance, the curves start from the corresponding sky-projected separation. We also show  dashed vertical lines at $D_{sep}\sqrt{3/2}$, representing the most likely 3D separations.

If DF4 and all three candidate hosts were at a similar distance of $D\approx 10$ Mpc rather than 20 Mpc, its MOND-predicted velocity dispersion would be $\sqrt{2} \times$ lower. As a result, the isolated velocity dispersion in the deep-MOND limit would fall from 18.6 km/s to 13.2 km/s in the complete absence of the EFE. Of course, some external field may be present if the object is at $D\approx 10\,$Mpc.

\subsubsection{NGC 1035}
\label{NGC1035}

We assume that NGC 1035 and DF4 are at similar distances of $D{\approx 20}$ Mpc, consistent with the Type \rm{II} supernova distance to the former of ${22 \pm 3}$ Mpc \citep{Schmidt_1992} or ${18 \pm 3}$ Mpc \citep{Schmidt_1994, Poznanski_2009}. The Tully-Fisher distance is smaller \citep[${14 \pm 3}$ Mpc,][]{Sorce_2014}, but this can be understood if the Tully-Fisher relation \citep{Tully_Fisher_1977} is a consequence of MOND for \emph{isolated} galaxies \citep{Milgrom83b}. Given the rather low surface brightness of NGC 1035 and the nearby massive galaxies NGC 1042 and NGC 1052, their external fields could reduce the circular velocity of NGC 1035 \citep{Haghi16}. If the EFE is not accounted for, a 10\% reduction in the circular velocity $v_f$ implies that the MOND dynamical mass must be 40\% lower. Assuming a fixed $M/L$, this is possible only for a 20\% lower distance, sufficient to explain why the Tully-Fisher distance to NGC 1035 is smaller than the Type II supernova distance by about this amount. The available information thus suggests NGC 1035 to lie at $D\approx 20$ Mpc.

Since DF4's sky-projected distance from NGC 1035 is only 21 kpc (assuming $D=20$ Mpc), its EFE on DF4 might significantly lower the latter's internal accelerations (top panel of Figure \ref{DF4}). Although such a small separation is discouraged by lack of tidal features around NGC 1035 and DF4 \citep{Muller_2019_image}, the two galaxies could plausibly be separated by ${\approx 100}$ kpc. This in turn somewhat reduces the MOND expectation for $\sigma_{intr}$.

If DF4 and NGC 1035 are close to each other, then tidal stability of DF4 could be an issue (see Sec. \ref{NGC1035_tides}). For the moment, we simply mention that at a separation of 60 kpc and distance of 20 Mpc, the EFE from NGC 1035 alone is sufficient to bring the MOND-predicted velocity dispersion of DF4 in agreement with the observed value \citep{vD19_DF4} at the 2-sigma confidence level (top panel of Figure \ref{DF4}). Because a lower distance reduces the predicted velocity dispersion, this improves to 1-sigma agreement if both objects are only 10 Mpc from us and 30 kpc from each other.

\subsubsection{NGC 1052}

Even if NGC 1035 is nowhere near DF4, its observed velocity dispersion is still rather sensitive to other possible hosts due to the low internal acceleration of DF4. In fact, its internal velocity dispersion is consistent with MOND at the 2-sigma confidence level once we consider the effect of NGC 1052 due to its high mass of $M_{\rm NGC1052}=10^{11}\,M_\odot$ at a projected separation of 167 kpc for $D = 20$ Mpc. This  improves to a 1$\sigma$ agreement if both NGC 1052 and DF4 are only $D = 10$ Mpc away and as close as possible to each other.

\subsubsection{NGC 1042}

The baryonic mass of NGC 1042 is $M_{\rm NGC1042}=2.4\times 10^{10}\,M_\odot$, roughly ${5 \times}$ less than that of NGC 1052. As a result, NGC 1042 has only a small effect on the internal dynamics of DF4, even if they have no line of sight separation (middle panel of Figure \ref{DF4}).

\subsubsection{Combined effect of multiple hosts}
\label{Combined_EFE}

Considering external fields from all these galaxies (assuming they are all at a similar distance $D$ from Earth and form a galaxy group) would perhaps lead to a lower $\sigma$ than when considering e.g. NGC 1035 alone. As NGC 1035, NGC 1042 and NGC 1052 are all in a similar direction from DF4, the external fields would add, making this a reasonable approximation in an upcoming project with full MONDian $N$-body simulations.

%If one additionally considers NGC 1035 which has a much smaller angular separation with DF4, then the MOND value is clearly in agreement within the 1-$\sigma$ confidence range of the observed velocity dispersion \citep{vD19_DF4} for a distance of either 10 or 20 Mpc. Even if $D_{sep}=60$ kpc relative to NGC 1035, the expected MOND velocity dispersion of DF4 still agrees with observations within the 1$\sigma$ (2$\sigma$) confidence level if the galaxies are at downrange distances of $D\approx 10$ ($D\approx 20$) Mpc.

\subsection{Tides from NGC 1035}
\label{NGC1035_tides}

The tidal radius of DF4 would be rather small if it was indeed only 21 kpc from NGC 1035, which is the minimum consistent with their observed angular separation for a distance of $D=20$ Mpc. In this case, DF4 would be in the process of tidal disruption, contradicting deep imaging data \citep[][figure 3]{Muller_2019_image}. The MONDian tidal radius of a mass $m$ located at separation $D_{sep}$ from another object of mass $M \gg m$ is \citep[][equation 14]{Zhao_2005}
\begin{eqnarray}
    r_t ~=~ 0.374 D \sqrt[3]{m/M} \, .
\end{eqnarray}

Given the total baryonic mass of NGC 1035, the minimum tidal radius of DF4 would be about 1.7 kpc assuming a separation of $D_{sep}=21$ kpc. This is comparable to its observed half-light radius of ${R}_{{\rm{e}}}=1.6\,\mathrm{kpc}$ \citep{vD19_DF4}. A larger separation distance ${\geq 100}$ kpc from NGC 1035 leads to a larger tidal radius ${\geq 8}$ kpc for DF4 such that tides do not affect it very much.% Even so, the EFE from other galaxies can still reduce the internal velocity dispersion of DF4 by a few km/s, enough to make it consistent with observations (middle panel of Fig. \ref{DF4}).

Tides from NGC 1035 would affect DF4 much less if DF4 is at a different distance than NGC 1035. In this case, neither the tidal nor the external field of NGC 1035 would affect DF4. However, our results in this section demonstrate that the EFE from NGC 1052 alone is sufficient to bring $\sigma_{\rm M, EF}$ in line with observations even if the distance to both is $D=20$ Mpc. The agreement improves further if both DF4 and NGC 1052 are closer to Earth, as long as they are also close to each other. At 10 Mpc, even an isolated DF4 is consistent with MOND at 2$\sigma$. The only problematic case is an isolated DF4 at 20 Mpc, which is just outside the 3-sigma observational upper limit (Figure \ref{DF4}). % A reduced distance from Earth is suggested by the fact that \citet{van_Dokkum_DF2_distance} obtained a surface brightness fluctuation distance of ${\approx 20}$ Mpc to DF2, similar to the value reported for DF4 \citep{vD19_DF4}. More detailed Hubble Space Telescope-based measurements of DF2 later indicated that it is very likely only 13 Mpc away based on direct detection of the tip of its red giant branch \citep{Trujillo2018}. As discussed in Section \ref{Distance_implications_DF2}, NGC 1052 could also be at a distance of 13 Mpc and thus impose a significant EFE on DF4, as assumed in Figure \ref{DF4}.

\section{Conclusions} \label{sec:conclusion}

Using previously conducted $N$-body simulations, we develop a fully analytical formulation of the MOND external field effect. We use this to calculate the velocity dispersion of the GC system of DF2, which we predict to be $14\,$km/s if MOND is correct. Our analysis of the ten (eleven) observed line of sight velocities of its GCs shows this prediction to be consistent with observations. Our analytical external field effect calculation agrees well with an independent estimation of the MONDian velocity dispersion \citep{FM12}. We test our analytical results using the first fully self-consistent PoR \citep{Lueghausen14b} $N$-body models of satellite galaxies orbiting a live host (Section \ref{sec:Nbody}). These concur with our analytical formulation and suggest that DF2 may be in a deep freeze state, with an even lower velocity dispersion than calculated analytically \citep{Haghi09, WK13a}.
%IB: above paragraph rephrased slightly.

Before this can be viewed as a confirmation of MOND, the distance of DF2 is addressed critically (Section \ref{sec:Distance}). While it cannot be excluded that DF2 is at the nominal distance of the NGC 1052 group ($D\approx 20$ Mpc), it is found that it may also be at around half this distance. If this were the case, then DF2 would be a normal dwarf galaxy consistent with MOND and it may even be a normal dSph satellite galaxy.  We note here that NGC 1052 may itself be at $D\approx 13$ Mpc (see Sec. \ref{Distance_implications_DF2})

% of NGC 1042.}
%IB_OM: deleted of NGC 1042.

It is also worth noting that the analysis by \citet{vanDokkum+18c} adopted a very small $\sigma = 3.2\,$km/s instead of the  value of $\approx 8.0$ km/s inferred here from their original data. By including a revised velocity for one of the GCs, \citet{vanDokkum+18b} later corrected the velocity dispersion to $\sigma=7.8^{+5.2}_{-2.2}\,$km/s, in better agreement with the stellar velocity dispersion \citep{Emsellem_2018} and also with MOND \citep{vanDokkum+18b}. Moreover, \citet{vanDokkum+18c} adopted a high stellar population mass-to-light ratio of $M_*/L_{\rm V}=2$ rather than the average value typical for such systems \citep[$M/L_{\rm V}=1.6$, see fig. 9 of][]{Dab16a}. The distance $D\approx 20\,$Mpc adopted by \citet{vanDokkum+18c} may also seem high, given that DF2 becomes a highly unusual galaxy on grounds unrelated to the correct law of gravity. All these choices push the results towards less dark matter and tension with MOND. Here, we have shown that by taking the data at face value and a more conservative theoretical approach, DF2 is consistent with a central MOND prediction, namely the EFE \citep{Milgrom1986}.

%Most recently, studies have been carried out on the distance issue, confirming the previously indicated distance of 20 Mpc \citep{Blakeslee2018} and \citep{Fensch_2018} or at the opposite side claiming  a much shorter distance of 13 Mpc \citep{Trujillo2018}.
%IB_OM: above paragraph is highly confrontational, but probably the best way to deal with dishonest researchers like van Dokkum.

Future observations will need to ascertain if this galaxy is indeed at $D\approx 20\,$Mpc and how isolated it is. The null detection of gas in DF2 \citep{Sardone_2019, Chowdhury2018} suggests that it resides in a group environment and thus feels a significant external field, independently of the assumed gravity law. Our investigation of the DF2 GCs suggests a normal specific GC frequency with a normal size and brightness, if it lies at a distance of 13 Mpc. A recently claimed detection of the tip of its red giant branch does indeed yield a distance of $13.4 \pm 1.1$ Mpc \citep[section 4.1 in][]{Trujillo2018}. If the whole NGC 1052 group is at 13 Mpc instead of 20 Mpc, the MOND predicted velocity dispersions should be reduced by 20\%, making them more consistent with observations (Section \ref{Distance_implications_DF2}). In this case, the high peculiar velocities of DF2 and of NGC 1052 relative to the CMB are in tension with the standard $\Lambda$CDM cosmological model but are well consistent with the velocity field expected in a MONDian universe \citep[figure 14 of][]{Candlish_2016}.

In Section \ref{sec:DF4}, we apply our analytic formalism to the recently discovered DF4 \citep{vD19_DF4}. Our analysis shows that the EFE from NGC 1052 could significantly reduce its MOND-predicted $\sigma$. Given their sky-projected separation, the effect could be strong enough to yield consistency with the observed $\sigma$ of DF4 (Figure \ref{DF4}). Even better agreement might be reached if one also considers the EFE it experiences from NGC 1035. We note that a lower distance than 20 Mpc further improves the agreement but the EFE from NGC 1042 can have only a small impact (Figure \ref{DF4}).  DF2 and DF4 would falsify MOND if these objects are completely isolated. In particular, the $8.5^{+2.3}_{-3.1}$ km/s velocity dispersion of DF2 reported by \citet{Danieli_2019} would rule out MOND at 3$\sigma$ (5$\sigma$) if it lies 13 Mpc (20 Mpc) from Earth.
%IB_OM: second-last sentence adjusted.

In a $\Lambda$CDM context, the rather low velocity dispersions of DF2 and DF4 suggest that they might be DM-poor tidal dwarf galaxies. In this regard, it is interesting to note that their radial velocities have opposite signs, once the systemic motion of NGC 1052 is subtracted. Thus, they may both be ancient metal-poor tidal dwarf galaxies orbiting NGC 1052 \citep{Recchi15}. This is reminiscent of the results obtained by \citet{Ibata14b}. The predicted existence of dark-matter-free tidal dwarf galaxies in a $\Lambda$CDM universe has been demonstrated conclusively by \citet{Haslbauer_2019}.

%In Section \ref{Satellite_plane}, we consider the possibility that the DF galaxies are distributed in a planar arrangement, similarly to the Local Group and Centaurus A satellite planes. At present, there is too little information to judge the validity of this hypothesis. Even so, it is worth considering whether an anisotropic arrangement eventually becomes evident. If it does and if the position along the long axis is strongly correlated with the radial velocity, then this would suggest a coherently rotating structure. This might be a distant satellite galaxy plane orbiting around NGC 1042 or NGC 1052. Future observations will shed more light on whether this is a viable idea.
%IB_OM: paragraph deleted. This is the way to deal with suggestions disproved by imaging.

Although DF2 and DF4 seem to contradict MOND at first glance, their velocity dispersions are actually well consistent with MOND expectations once the EFE is included. The EFE is an integral part of MOND that follows directly from its governing equations \citep{Milgrom1986}. Thus, careful analytical and numerical work is required to judge what MOND really predicts for any individual system. Although not a trivial task, this can in principle be done rather accurately because MOND relies only on the distribution of actually observed baryonic matter.

Finally, it is clear from this discussion that a critical unknown in our understanding of DF2 and DF4 and whether they are associated with NGC 1052 is the distance problem: just how far from us are these galaxies?

\section*{Acknowledgements}

AHZ and IB are Alexander von Humboldt Fellows. HH is a DAAD visiting scholar. BJ thanks the hospitality of the Stellar Populations and Dynamics Research Group in Bonn and of the AIfA, where this work was done. OM thanks the Swiss National Science Foundation for financial support. XW gives thanks for support from the Natural Science Foundation of China grants numbers 11503025 and 11421303, Anhui Natural Science Foundation grant number 1708085MA20 and the ``Hundred Talents Project of Anhui Province''.
%\end{acknowledgments}

\bibliographystyle{mnras} % Tell bibtex which bibliography style to use
%\bibliography{/Users/pavel/PAPERS/BIBL_REFERENCES/kroupa_ref.bib} %
\bibliography{ref}

\begin{thebibliography}{}
\makeatletter
\relax
\def\mn@urlcharsother{\let\do\@makeother \do\$\do\&\do\#\do\^\do\_\do\%\do\~}
\def\mn@doi{\begingroup\mn@urlcharsother \@ifnextchar [ {\mn@doi@}
  {\mn@doi@[]}}
\def\mn@doi@[#1]#2{\def\@tempa{#1}\ifx\@tempa\@empty \href
  {http://dx.doi.org/#2} {doi:#2}\else \href {http://dx.doi.org/#2} {#1}\fi
  \endgroup}
\def\mn@eprint#1#2{\mn@eprint@#1:#2::\@nil}
\def\mn@eprint@arXiv#1{\href {http://arxiv.org/abs/#1} {{\tt arXiv:#1}}}
\def\mn@eprint@dblp#1{\href {http://dblp.uni-trier.de/rec/bibtex/#1.xml}
  {dblp:#1}}
\def\mn@eprint@#1:#2:#3:#4\@nil{\def\@tempa {#1}\def\@tempb {#2}\def\@tempc
  {#3}\ifx \@tempc \@empty \let \@tempc \@tempb \let \@tempb \@tempa \fi \ifx
  \@tempb \@empty \def\@tempb {arXiv}\fi \@ifundefined
  {mn@eprint@\@tempb}{\@tempb:\@tempc}{\expandafter \expandafter \csname
  mn@eprint@\@tempb\endcsname \expandafter{\@tempc}}}

\bibitem[\protect\citeauthoryear{{Banik} \& {Zhao}}{{Banik} \&
  {Zhao}}{2018a}]{banik18}
{Banik} I.,  {Zhao} H.,  2018a, \mn@doi [MNRAS] {10.1093/mnras/stx2350}, \href
  {http://adsabs.harvard.edu/abs/2018MNRAS.473..419B} {473, 419}

\bibitem[\protect\citeauthoryear{{Banik} \& {Zhao}}{{Banik} \&
  {Zhao}}{2018b}]{BZ18}
{Banik} I.,  {Zhao} H.,  2018b, \mn@doi [MNRAS] {10.1093/mnras/stx2596}, \href
  {http://adsabs.harvard.edu/abs/2018MNRAS.473.4033B} {473, 4033}

\bibitem[\protect\citeauthoryear{{Beers}, {Flynn}  \& {Gebhardt}}{{Beers}
  et~al.}{1990}]{Beers+90}
{Beers} T.~C.,  {Flynn} K.,   {Gebhardt} K.,  1990, \mn@doi [\aj]
  {10.1086/115487}, \href {http://adsabs.harvard.edu/abs/1990AJ....100...32B}
  {100, 32}

\bibitem[\protect\citeauthoryear{{Bekenstein} \& {Milgrom}}{{Bekenstein} \&
  {Milgrom}}{1984}]{BM84}
{Bekenstein} J.,  {Milgrom} M.,  1984, \mn@doi [ApJ] {10.1086/162570}, \href
  {http://adsabs.harvard.edu/abs/1984ApJ...286....7B} {286, 7}

\bibitem[\protect\citeauthoryear{{Bellstedt} et~al.,}{{Bellstedt}
  et~al.}{2018}]{Bellstedt+18}
{Bellstedt} S.,  et~al., 2018, \mn@doi [MNRAS] {10.1093/mnras/sty456}, \href
  {https://ui.adsabs.harvard.edu/#abs/2018MNRAS.476.4543B} {476, 4543}

\bibitem[\protect\citeauthoryear{{Bose} et~al.,}{{Bose}
  et~al.}{2018}]{Bose_2018}
{Bose} S.,  et~al., 2018, preprint, \href
  {http://adsabs.harvard.edu/abs/2018arXiv181003635B} {Arxiv} (\mn@eprint
  {arXiv} {1810.03635})

\bibitem[\protect\citeauthoryear{{Brough}, {Forbes}, {Kilborn}  \&
  {Couch}}{{Brough} et~al.}{2006}]{Brough_2006}
{Brough} S.,  {Forbes} D.~A.,  {Kilborn} V.~A.,   {Couch} W.,  2006, \mn@doi
  [\mnras] {10.1111/j.1365-2966.2006.10542.x}, \href
  {https://ui.adsabs.harvard.edu/abs/2006MNRAS.370.1223B} {370, 1223}

\bibitem[\protect\citeauthoryear{{Bullock} \& {Boylan-Kolchin}}{{Bullock} \&
  {Boylan-Kolchin}}{2017}]{Bullock_2017}
{Bullock} J.~S.,  {Boylan-Kolchin} M.,  2017, \mn@doi [Annual Review of
  Astronomy and Astrophysics] {10.1146/annurev-astro-091916-055313}, \href
  {https://ui.adsabs.harvard.edu/abs/2017ARA\&A..55..343B} {55, 343}

\bibitem[\protect\citeauthoryear{{Caldwell} et~al.,}{{Caldwell}
  et~al.}{2017}]{Caldwell_2017}
{Caldwell} N.,  et~al., 2017, \mn@doi [ApJ] {10.3847/1538-4357/aa688e}, \href
  {http://adsabs.harvard.edu/abs/2017ApJ...839...20C} {839, 20}

\bibitem[\protect\citeauthoryear{{Candlish}}{{Candlish}}{2016}]{Candlish_2016}
{Candlish} G.~N.,  2016, \mn@doi [\mnras] {10.1093/mnras/stw1130}, \href
  {http://adsabs.harvard.edu/abs/2016MNRAS.460.2571C} {460, 2571}

\bibitem[\protect\citeauthoryear{{Chowdhury}}{{Chowdhury}}{2019}]{Chowdhury2018}
{Chowdhury} A.,  2019, \mn@doi [MNRAS] {10.1093/mnrasl/sly192}, \href
  {https://ui.adsabs.harvard.edu/#abs/2019MNRAS.482L..99C} {482, L99}

\bibitem[\protect\citeauthoryear{{Cole}, {Dehnen}, {Read}  \&
  {Wilkinson}}{{Cole} et~al.}{2012}]{Cole+12}
{Cole} D.~R.,  {Dehnen} W.,  {Read} J.~I.,   {Wilkinson} M.~I.,  2012, \mn@doi
  [MNRAS] {10.1111/j.1365-2966.2012.21885.x}, \href
  {http://adsabs.harvard.edu/abs/2012MNRAS.426..601C} {426, 601}

\bibitem[\protect\citeauthoryear{{Dabringhausen} \&
  {Fellhauer}}{{Dabringhausen} \& {Fellhauer}}{2016}]{Dab16a}
{Dabringhausen} J.,  {Fellhauer} M.,  2016, \mn@doi [MNRAS]
  {10.1093/mnras/stw1248}, \href
  {http://adsabs.harvard.edu/abs/2016MNRAS.460.4492D} {460, 4492}

\bibitem[\protect\citeauthoryear{{Danieli}, {van Dokkum}, {Conroy}, {Abraham}
  \& {Romanowsky}}{{Danieli} et~al.}{2019}]{Danieli_2019}
{Danieli} S.,  {van Dokkum} P.,  {Conroy} C.,  {Abraham} R.,   {Romanowsky}
  A.~J.,  2019, \mn@doi [ApJL] {10.3847/2041-8213/ab0e8c}, \href
  {https://ui.adsabs.harvard.edu/abs/2019ApJ...874L..12D} {874, L12}

\bibitem[\protect\citeauthoryear{{Di Cintio} \& {Lelli}}{{Di Cintio} \&
  {Lelli}}{2016}]{DiCintio_2016}
{Di Cintio} A.,  {Lelli} F.,  2016, \mn@doi [\mnras] {10.1093/mnrasl/slv185},
  \href {https://ui.adsabs.harvard.edu/abs/2016MNRAS.456L.127D} {456, L127}

\bibitem[\protect\citeauthoryear{{Elmegreen}}{{Elmegreen}}{1999}]{Elmegreen_1999}
{Elmegreen} B.~G.,  1999, \mn@doi [\apss] {10.1023/A:1017094323836}, \href
  {https://ui.adsabs.harvard.edu/#abs/1999Ap&SS.269..469E} {269, 469}

\bibitem[\protect\citeauthoryear{{Emsellem} et~al.,}{{Emsellem}
  et~al.}{2018}]{Emsellem_2018}
{Emsellem} E.,  et~al., 2018, \mn@doi [\aap] {10.1051/0004-6361/201834909},
  \href {https://ui.adsabs.harvard.edu/#abs/2018arXiv181207345E} {accepted}

\bibitem[\protect\citeauthoryear{{Famaey} \& {Binney}}{{Famaey} \&
  {Binney}}{2005}]{Famaey_2005}
{Famaey} B.,  {Binney} J.,  2005, \mn@doi [\mnras]
  {10.1111/j.1365-2966.2005.09474.x}, \href
  {http://adsabs.harvard.edu/abs/2005MNRAS.363..603F} {363, 603}

\bibitem[\protect\citeauthoryear{{Famaey} \& {McGaugh}}{{Famaey} \&
  {McGaugh}}{2012}]{FM12}
{Famaey} B.,  {McGaugh} S.~S.,  2012, \mn@doi [Living Reviews in Relativity]
  {10.12942/lrr-2012-10}, \href
  {http://adsabs.harvard.edu/abs/2012LRR....15...10F} {15, 10}

\bibitem[\protect\citeauthoryear{{Famaey}, {McGaugh}  \& {Milgrom}}{{Famaey}
  et~al.}{2018}]{Famaey18}
{Famaey} B.,  {McGaugh} S.,   {Milgrom} M.,  2018, \mn@doi [MNRAS]
  {10.1093/mnras/sty1884}, \href
  {http://adsabs.harvard.edu/abs/2018MNRAS.480..473F} {480, 473}

\bibitem[\protect\citeauthoryear{{Fosbury}, {Mebold}, {Goss}  \&
  {Dopita}}{{Fosbury} et~al.}{1978}]{Fosbury_1978}
{Fosbury} R.~A.~E.,  {Mebold} U.,  {Goss} W.~M.,   {Dopita} M.~A.,  1978,
  \mn@doi [MNRAS] {10.1093/mnras/183.4.549}, \href
  {http://adsabs.harvard.edu/abs/1978MNRAS.183..549F} {183, 549}

\bibitem[\protect\citeauthoryear{{Geha}, {Blanton}, {Masjedi}  \&
  {West}}{{Geha} et~al.}{2006}]{Geha06}
{Geha} M.,  {Blanton} M.~R.,  {Masjedi} M.,   {West} A.~A.,  2006, \mn@doi
  [ApJ] {10.1086/508604}, \href
  {http://adsabs.harvard.edu/abs/2006ApJ...653..240G} {653, 240}

\bibitem[\protect\citeauthoryear{{Gennaro} et~al.,}{{Gennaro}
  et~al.}{2018}]{Gennaro+18}
{Gennaro} M.,  et~al., 2018, \mn@doi [ApJ] {10.3847/1538-4357/aaa973}, \href
  {http://adsabs.harvard.edu/abs/2018ApJ...855...20G} {855, 20}

\bibitem[\protect\citeauthoryear{{Georgiev}, {Puzia}, {Goudfrooij}  \&
  {Hilker}}{{Georgiev} et~al.}{2010}]{Georgiev_2010}
{Georgiev} I.~Y.,  {Puzia} T.~H.,  {Goudfrooij} P.,   {Hilker} M.,  2010,
  \mn@doi [\mnras] {10.1111/j.1365-2966.2010.16802.x}, \href
  {http://adsabs.harvard.edu/abs/2010MNRAS.406.1967G} {406, 1967}

\bibitem[\protect\citeauthoryear{{Gunawardhana} et~al.,}{{Gunawardhana}
  et~al.}{2011}]{Gun11}
{Gunawardhana} M.~L.~P.,  et~al., 2011, \mn@doi [MNRAS]
  {10.1111/j.1365-2966.2011.18800.x}, \href
  {http://adsabs.harvard.edu/abs/2011MNRAS.415.1647G} {415, 1647}

\bibitem[\protect\citeauthoryear{{Haghi}, {Baumgardt}, {Kroupa}, {Grebel},
  {Hilker}  \& {Jordi}}{{Haghi} et~al.}{2009}]{Haghi09}
{Haghi} H.,  {Baumgardt} H.,  {Kroupa} P.,  {Grebel} E.~K.,  {Hilker} M.,
  {Jordi} K.,  2009, \mn@doi [MNRAS] {10.1111/j.1365-2966.2009.14656.x}, \href
  {http://adsabs.harvard.edu/abs/2009MNRAS.395.1549H} {395, 1549}

\bibitem[\protect\citeauthoryear{{Haghi}, {Bazkiaei}, {Zonoozi}  \&
  {Kroupa}}{{Haghi} et~al.}{2016}]{Haghi16}
{Haghi} H.,  {Bazkiaei} A.~E.,  {Zonoozi} A.~H.,   {Kroupa} P.,  2016, \mn@doi
  [MNRAS] {10.1093/mnras/stw573}, \href
  {http://adsabs.harvard.edu/abs/2016MNRAS.458.4172H} {458, 4172}

\bibitem[\protect\citeauthoryear{{Haslbauer}, {Dabringhausen}, {Kroupa},
  {Javanmardi}  \& {Banik}}{{Haslbauer} et~al.}{2019}]{Haslbauer_2019}
{Haslbauer} M.,  {Dabringhausen} J.,  {Kroupa} P.,  {Javanmardi} B.,   {Banik}
  I.,  2019, \mn@doi [\aap] {10.1051/0004-6361/201833771}, \href
  {https://ui.adsabs.harvard.edu/abs/2019arXiv190503258H} {accepted}

\bibitem[\protect\citeauthoryear{{Hees}, {Famaey}, {Angus}  \&
  {Gentile}}{{Hees} et~al.}{2016}]{Hees16}
{Hees} A.,  {Famaey} B.,  {Angus} G.~W.,   {Gentile} G.,  2016, \mn@doi [MNRAS]
  {10.1093/mnras/stv2330}, \href
  {http://adsabs.harvard.edu/abs/2016MNRAS.455..449H} {455, 449}

\bibitem[\protect\citeauthoryear{{Hernquist}}{{Hernquist}}{1990}]{Hernquist1991}
{Hernquist} L.,  1990, \mn@doi [\apj] {10.1086/168845}, \href
  {http://adsabs.harvard.edu/abs/1990ApJ...356..359H} {356, 359}

\bibitem[\protect\citeauthoryear{{Hoof}, {Geringer-Sameth}  \& {Trotta}}{{Hoof}
  et~al.}{2018}]{Hoof_2019}
{Hoof} S.,  {Geringer-Sameth} A.,   {Trotta} R.,  2018, preprint, \href
  {http://adsabs.harvard.edu/abs/2018arXiv181206986H} {Arxiv} (\mn@eprint
  {arXiv} {1812.06986})

\bibitem[\protect\citeauthoryear{{Ibata}, {Ibata}, {Famaey}  \&
  {Lewis}}{{Ibata} et~al.}{2014}]{Ibata14b}
{Ibata} N.~G.,  {Ibata} R.~A.,  {Famaey} B.,   {Lewis} G.~F.,  2014, \mn@doi
  [\nat] {10.1038/nature13481}, \href
  {http://adsabs.harvard.edu/abs/2014Natur.511..563I} {511, 563}

\bibitem[\protect\citeauthoryear{{Jerjen} \& {Rejkuba}}{{Jerjen} \&
  {Rejkuba}}{2001}]{JR01}
{Jerjen} H.,  {Rejkuba} M.,  2001, \mn@doi [\aap] {10.1051/0004-6361:20010389},
  \href {http://adsabs.harvard.edu/abs/2001A\%26A...371..487J} {371, 487}

\bibitem[\protect\citeauthoryear{{Je{\v r}{\'a}bkov{\'a}}, {Hasani Zonoozi},
  {Kroupa}, {Beccari}, {Yan}, {Vazdekis}  \& {Zhang}}{{Je{\v r}{\'a}bkov{\'a}}
  et~al.}{2018}]{Jerabkova18}
{Je{\v r}{\'a}bkov{\'a}} T.,  {Hasani Zonoozi} A.,  {Kroupa} P.,  {Beccari} G.,
   {Yan} Z.,  {Vazdekis} A.,   {Zhang} Z.-Y.,  2018, \mn@doi [\aap]
  {10.1051/0004-6361/201833055}, \href
  {http://adsabs.harvard.edu/abs/2018A\%26A...620A..39J} {620, A39}

\bibitem[\protect\citeauthoryear{{Kogut} et~al.,}{{Kogut}
  et~al.}{1993}]{Kogut93}
{Kogut} A.,  et~al., 1993, \mn@doi [ApJ] {10.1086/173453}, \href
  {http://adsabs.harvard.edu/abs/1993ApJ...419....1K} {419, 1}

\bibitem[\protect\citeauthoryear{{Kroupa}}{{Kroupa}}{1997}]{Kroupa97}
{Kroupa} P.,  1997, \mn@doi [New Astronomy] {10.1016/S1384-1076(97)00012-2},
  \href {http://adsabs.harvard.edu/abs/1997NewA....2..139K} {2, 139}

\bibitem[\protect\citeauthoryear{{Kroupa}}{{Kroupa}}{2012}]{Kroupa12a}
{Kroupa} P.,  2012, \mn@doi [Publications of the Astronomical Society of
  Australia] {10.1071/AS12005}, \href
  {http://adsabs.harvard.edu/abs/2012PASA...29..395K} {29, 395}

\bibitem[\protect\citeauthoryear{{Kroupa}}{{Kroupa}}{2015}]{Kroupa15}
{Kroupa} P.,  2015, \mn@doi [Canadian Journal of Physics]
  {10.1139/cjp-2014-0179}, \href
  {http://adsabs.harvard.edu/abs/2015CaJPh..93..169K} {93, 169}

\bibitem[\protect\citeauthoryear{{Kroupa} et~al.,}{{Kroupa}
  et~al.}{2010}]{Kroupa10}
{Kroupa} P.,  et~al., 2010, \mn@doi [\aap] {10.1051/0004-6361/201014892}, \href
  {http://adsabs.harvard.edu/abs/2010A\%26A...523A..32K} {523, A32}

\bibitem[\protect\citeauthoryear{{Kroupa}, {Weidner}, {Pflamm-Altenburg},
  {Thies}, {Dabringhausen}, {Marks}  \& {Maschberger}}{{Kroupa}
  et~al.}{2013}]{Kroupa13}
{Kroupa} P.,  {Weidner} C.,  {Pflamm-Altenburg} J.,  {Thies} I.,
  {Dabringhausen} J.,  {Marks} M.,   {Maschberger} T.,  2013, {The Stellar and
  Sub-Stellar Initial Mass Function of Simple and Composite Populations}.
{Springer Science} and {Business Media Dordrecht}, p.~115,
  \mn@doi{10.1007/978-94-007-5612-0_4}

\bibitem[\protect\citeauthoryear{{Kroupa} et~al.,}{{Kroupa}
  et~al.}{2018}]{Kroupa_2018}
{Kroupa} P.,  et~al., 2018, \mn@doi [\nat] {10.1038/s41586-018-0429-z}, \href
  {http://adsabs.harvard.edu/abs/2018Natur.561E....4} {561, E4}

\bibitem[\protect\citeauthoryear{{Laporte}, {Agnello}  \& {Navarro}}{{Laporte}
  et~al.}{2019}]{Laporte_2019}
{Laporte} C.~F.~P.,  {Agnello} A.,   {Navarro} J.~F.,  2019, \mn@doi [\mnras]
  {10.1093/mnras/sty2891}, \href
  {http://adsabs.harvard.edu/abs/2019MNRAS.484..245L} {484, 245}

\bibitem[\protect\citeauthoryear{{Lee} et~al.,}{{Lee} et~al.}{2009}]{Lee09}
{Lee} J.~C.,  et~al., 2009, \mn@doi [ApJ] {10.1088/0004-637X/706/1/599}, \href
  {http://adsabs.harvard.edu/abs/2009ApJ...706..599L} {706, 599}

\bibitem[\protect\citeauthoryear{{Londrillo} \& {Nipoti}}{{Londrillo} \&
  {Nipoti}}{2009}]{NMODY}
{Londrillo} P.,  {Nipoti} C.,  2009, Memorie della Societa Astronomica Italiana
  Supplementi, \href {http://adsabs.harvard.edu/abs/2009MSAIS..13...89L} {13,
  89}

\bibitem[\protect\citeauthoryear{{L{\"u}ghausen}, {Famaey}  \&
  {Kroupa}}{{L{\"u}ghausen} et~al.}{2015}]{Lueghausen14b}
{L{\"u}ghausen} F.,  {Famaey} B.,   {Kroupa} P.,  2015, Canadian Journal of
  Physics, \href {http://adsabs.harvard.edu/abs/2014arXiv1405.5963L} {93, 232}

\bibitem[\protect\citeauthoryear{{Martin}, {de Jong}  \& {Rix}}{{Martin}
  et~al.}{2008}]{Martin_2008}
{Martin} N.~F.,  {de Jong} J. T.~A.,   {Rix} H.-W.,  2008, \mn@doi [\apj]
  {10.1086/590336}, \href
  {https://ui.adsabs.harvard.edu/abs/2008ApJ...684.1075M} {684, 1075}

\bibitem[\protect\citeauthoryear{{Martin}, {Collins}, {Longeard}  \&
  {Tollerud}}{{Martin} et~al.}{2018}]{Martin_2018}
{Martin} N.~F.,  {Collins} M. L.~M.,  {Longeard} N.,   {Tollerud} E.,  2018,
  \mn@doi [\apj] {10.3847/2041-8213/aac216}, \href
  {https://ui.adsabs.harvard.edu/#abs/2018ApJ...859L...5M} {859, L5}

\bibitem[\protect\citeauthoryear{{Mateo}}{{Mateo}}{1998}]{Mateo_1998}
{Mateo} M.~L.,  1998, \mn@doi [\araa] {10.1146/annurev.astro.36.1.435}, \href
  {http://adsabs.harvard.edu/abs/1998ARA\%26A..36..435M} {36, 435}

\bibitem[\protect\citeauthoryear{{McConnachie}}{{McConnachie}}{2012}]{McConnachie12}
{McConnachie} A.~W.,  2012, \mn@doi [\aj] {10.1088/0004-6256/144/1/4}, \href
  {http://adsabs.harvard.edu/abs/2012AJ....144....4M} {144, 4}

\bibitem[\protect\citeauthoryear{{McGaugh}}{{McGaugh}}{2016}]{McGaugh16a}
{McGaugh} S.~S.,  2016, \mn@doi [ApJl] {10.3847/2041-8205/832/1/L8}, \href
  {http://adsabs.harvard.edu/abs/2016ApJ...832L...8M} {832, L8}

\bibitem[\protect\citeauthoryear{{McGaugh} \& {Milgrom}}{{McGaugh} \&
  {Milgrom}}{2013}]{MM13}
{McGaugh} S.,  {Milgrom} M.,  2013, \mn@doi [ApJ]
  {10.1088/0004-637X/775/2/139}, \href
  {http://adsabs.harvard.edu/abs/2013ApJ...775..139M} {775, 139}

\bibitem[\protect\citeauthoryear{{McGaugh}, {Schombert}, {Bothun}  \& {de
  Blok}}{{McGaugh} et~al.}{2000}]{McGaugh2000}
{McGaugh} S.~S.,  {Schombert} J.~M.,  {Bothun} G.~D.,   {de Blok} W.~J.~G.,
  2000, \mn@doi [\apjl] {10.1086/312628}, \href
  {http://adsabs.harvard.edu/abs/2000ApJ...533L..99M} {533, L99}

\bibitem[\protect\citeauthoryear{{Milgrom}}{{Milgrom}}{1983a}]{Milgrom83}
{Milgrom} M.,  1983a, \mn@doi [ApJ] {10.1086/161130}, \href
  {http://adsabs.harvard.edu/abs/1983ApJ...270..365M} {270, 365}

\bibitem[\protect\citeauthoryear{{Milgrom}}{{Milgrom}}{1983b}]{Milgrom83b}
{Milgrom} M.,  1983b, \mn@doi [\apj] {10.1086/161131}, \href
  {http://adsabs.harvard.edu/abs/1983ApJ...270..371M} {270, 371}

\bibitem[\protect\citeauthoryear{{Milgrom}}{{Milgrom}}{1986}]{Milgrom1986}
{Milgrom} M.,  1986, \mn@doi [ApJ] {10.1086/164021}, \href
  {http://adsabs.harvard.edu/abs/1986ApJ...302..617M} {302, 617}

\bibitem[\protect\citeauthoryear{{Milgrom}}{{Milgrom}}{1994}]{Milgrom94}
{Milgrom} M.,  1994, \mn@doi [ApJ] {10.1086/174341}, \href
  {http://adsabs.harvard.edu/abs/1994ApJ...429..540M} {429, 540}

\bibitem[\protect\citeauthoryear{{Milgrom}}{{Milgrom}}{1995}]{Milgrom_1995}
{Milgrom} M.,  1995, \mn@doi [\apj] {10.1086/176592}, \href
  {http://adsabs.harvard.edu/abs/1995ApJ...455..439M} {455, 439}

\bibitem[\protect\citeauthoryear{{Milgrom}}{{Milgrom}}{2009}]{Milgrom09}
{Milgrom} M.,  2009, \mn@doi [ApJ] {10.1088/0004-637X/698/2/1630}, \href
  {http://adsabs.harvard.edu/abs/2009ApJ...698.1630M} {698, 1630}

\bibitem[\protect\citeauthoryear{{M{\"u}ller}, {Jerjen}  \&
  {Binggeli}}{{M{\"u}ller} et~al.}{2018}]{Muller18}
{M{\"u}ller} O.,  {Jerjen} H.,   {Binggeli} B.,  2018, \mn@doi [\aap]
  {10.1051/0004-6361/201832897}, \href
  {http://adsabs.harvard.edu/abs/2018A\%26A...615A.105M} {615, A105}

\bibitem[\protect\citeauthoryear{{M{\"u}ller}, {Famaey}  \&
  {Zhao}}{{M{\"u}ller} et~al.}{2019a}]{Muller_2019}
{M{\"u}ller} O.,  {Famaey} B.,   {Zhao} H.,  2019a, \mn@doi [\aap]
  {10.1051/0004-6361/201834914}, \href
  {http://adsabs.harvard.edu/abs/2019A\%26A...623A..36M} {623, A36}

\bibitem[\protect\citeauthoryear{{M{\"u}ller} et~al.,}{{M{\"u}ller}
  et~al.}{2019b}]{Muller_2019_image}
{M{\"u}ller} O.,  et~al., 2019b, \mn@doi [A\&A Letters]
  {10.1051/0004-6361/201935463}, \href
  {http://adsabs.harvard.edu/abs/2019arXiv190307285M} {accepted}

\bibitem[\protect\citeauthoryear{{Ostriker} \& {Peebles}}{{Ostriker} \&
  {Peebles}}{1973}]{OP73}
{Ostriker} J.~P.,  {Peebles} P.~J.~E.,  1973, \mn@doi [ApJ] {10.1086/152513},
  \href {http://adsabs.harvard.edu/abs/1973ApJ...186..467O} {186, 467}

\bibitem[\protect\citeauthoryear{{Ostriker} \& {Steinhardt}}{{Ostriker} \&
  {Steinhardt}}{1995}]{Ostriker_Steinhardt_1995}
{Ostriker} J.~P.,  {Steinhardt} P.~J.,  1995, \mn@doi [Nature]
  {10.1038/377600a0}, \href {http://adsabs.harvard.edu/abs/1995Natur.377..600O}
  {377, 600}

\bibitem[\protect\citeauthoryear{{Pawlowski} \& {McGaugh}}{{Pawlowski} \&
  {McGaugh}}{2014}]{PM14}
{Pawlowski} M.~S.,  {McGaugh} S.~S.,  2014, \mn@doi [MNRAS]
  {10.1093/mnras/stu321}, \href
  {http://adsabs.harvard.edu/abs/2014MNRAS.440..908P} {440, 908}

\bibitem[\protect\citeauthoryear{{Peebles} \& {Nusser}}{{Peebles} \&
  {Nusser}}{2010}]{PN10}
{Peebles} P.~J.~E.,  {Nusser} A.,  2010, \mn@doi [\nat] {10.1038/nature09101},
  \href {http://adsabs.harvard.edu/abs/2010Natur.465..565P} {465, 565}

\bibitem[\protect\citeauthoryear{{Planck Collaboration}}{{Planck
  Collaboration}}{2016}]{Planck16}
{Planck Collaboration} 2016, \mn@doi [\aap] {10.1051/0004-6361/201525830},
  \href {http://adsabs.harvard.edu/abs/2016A\%26A...594A..13P} {594, A13}

\bibitem[\protect\citeauthoryear{{Poznanski} et~al.,}{{Poznanski}
  et~al.}{2009}]{Poznanski_2009}
{Poznanski} D.,  et~al., 2009, \mn@doi [\apj] {10.1088/0004-637X/694/2/1067},
  \href {http://adsabs.harvard.edu/abs/2009ApJ...694.1067P} {694, 1067}

\bibitem[\protect\citeauthoryear{{Recchi} \& {Kroupa}}{{Recchi} \&
  {Kroupa}}{2015}]{Recchi15}
{Recchi} S.,  {Kroupa} P.,  2015, \mn@doi [MNRAS] {10.1093/mnras/stu2338},
  \href {http://adsabs.harvard.edu/abs/2015MNRAS.446.4168R} {446, 4168}

\bibitem[\protect\citeauthoryear{{Rejkuba}}{{Rejkuba}}{2012}]{Rejkuba12}
{Rejkuba} M.,  2012, \mn@doi [\apss] {10.1007/s10509-012-0986-9}, \href
  {http://adsabs.harvard.edu/abs/2012Ap\%26SS.341..195R} {341, 195}

\bibitem[\protect\citeauthoryear{{Sardone}, {Pisano}, {Burke-Spolaor},
  {Mascoop}  \& {Pol}}{{Sardone} et~al.}{2019}]{Sardone_2019}
{Sardone} A.,  {Pisano} D.~J.,  {Burke-Spolaor} S.,  {Mascoop} J.~L.,   {Pol}
  N.,  2019, \mn@doi [ApJL] {10.3847/2041-8213/ab0084}, \href
  {http://adsabs.harvard.edu/abs/2019arXiv190107586S} {871, L31}

\bibitem[\protect\citeauthoryear{{Schmidt}, {Kirshner}  \& {Eastman}}{{Schmidt}
  et~al.}{1992}]{Schmidt_1992}
{Schmidt} B.~P.,  {Kirshner} R.~P.,   {Eastman} R.~G.,  1992, \mn@doi [\apj]
  {10.1086/171659}, \href {http://adsabs.harvard.edu/abs/1992ApJ...395..366S}
  {395, 366}

\bibitem[\protect\citeauthoryear{{Schmidt}, {Kirshner}, {Eastman}, {Phillips},
  {Suntzeff}, {Hamuy}, {Maza}  \& {Aviles}}{{Schmidt}
  et~al.}{1994}]{Schmidt_1994}
{Schmidt} B.~P.,  {Kirshner} R.~P.,  {Eastman} R.~G.,  {Phillips} M.~M.,
  {Suntzeff} N.~B.,  {Hamuy} M.,  {Maza} J.,   {Aviles} R.,  1994, \mn@doi
  [\apj] {10.1086/174546}, \href
  {http://adsabs.harvard.edu/abs/1994ApJ...432...42S} {432, 42}

\bibitem[\protect\citeauthoryear{{Skrutskie} et~al.,}{{Skrutskie}
  et~al.}{2006}]{Skrutskie_2006}
{Skrutskie} M.~F.,  et~al., 2006, \mn@doi [\aj] {10.1086/498708}, \href
  {https://ui.adsabs.harvard.edu/abs/2006AJ....131.1163S} {131, 1163}

\bibitem[\protect\citeauthoryear{{Sorce}, {Tully}, {Courtois}, {Jarrett},
  {Neill}  \& {Shaya}}{{Sorce} et~al.}{2014}]{Sorce_2014}
{Sorce} J.~G.,  {Tully} R.~B.,  {Courtois} H.~M.,  {Jarrett} T.~H.,  {Neill}
  J.~D.,   {Shaya} E.~J.,  2014, \mn@doi [\mnras] {10.1093/mnras/stu1450},
  \href {http://adsabs.harvard.edu/abs/2014MNRAS.444..527S} {444, 527}

\bibitem[\protect\citeauthoryear{{Theureau}, {Hanski}, {Coudreau}, {Hallet}  \&
  {Martin}}{{Theureau} et~al.}{2007}]{Theureau_2007}
{Theureau} G.,  {Hanski} M.~O.,  {Coudreau} N.,  {Hallet} N.,   {Martin} J.-M.,
   2007, \mn@doi [\aap] {10.1051/0004-6361:20066187}, \href
  {http://adsabs.harvard.edu/abs/2007A\%26A...465...71T} {465, 71}

\bibitem[\protect\citeauthoryear{{Thomas}, {Famaey}, {Ibata}, {Renaud},
  {Martin}  \& {Kroupa}}{{Thomas} et~al.}{2018}]{Thomas+18}
{Thomas} G.~F.,  {Famaey} B.,  {Ibata} R.,  {Renaud} F.,  {Martin} N.~F.,
  {Kroupa} P.,  2018, \mn@doi [\aap] {10.1051/0004-6361/201731609}, \href
  {http://adsabs.harvard.edu/abs/2018A\%26A...609A..44T} {609, A44}

\bibitem[\protect\citeauthoryear{{Trujillo} et~al.,}{{Trujillo}
  et~al.}{2019}]{Trujillo2018}
{Trujillo} I.,  et~al., 2019, \mn@doi [\mnras] {10.1093/mnras/stz771}, \href
  {http://adsabs.harvard.edu/abs/2019MNRAS.tmp..733T} {accepted}

\bibitem[\protect\citeauthoryear{{Tully} \& {Fisher}}{{Tully} \&
  {Fisher}}{1977}]{Tully_Fisher_1977}
{Tully} R.~B.,  {Fisher} J.~R.,  1977, \aap, \href
  {http://adsabs.harvard.edu/abs/1977A\%26A....54..661T} {54, 661}

\bibitem[\protect\citeauthoryear{{Watts}, {Meurer}, {Lagos}, {Bruzzese},
  {Kroupa}  \& {Jerabkova}}{{Watts} et~al.}{2018}]{Watts2018}
{Watts} A.~B.,  {Meurer} G.~R.,  {Lagos} C.~D.~P.,  {Bruzzese} S.~M.,  {Kroupa}
  P.,   {Jerabkova} T.,  2018, \mn@doi [MNRAS] {10.1093/mnras/sty1006}, \href
  {http://adsabs.harvard.edu/abs/2018MNRAS.477.5554W} {477, 5554}

\bibitem[\protect\citeauthoryear{{Wolf}, {Martinez}, {Bullock}, {Kaplinghat},
  {Geha}, {Mu{\~n}oz}, {Simon}  \& {Avedo}}{{Wolf} et~al.}{2010}]{Wolf10}
{Wolf} J.,  {Martinez} G.~D.,  {Bullock} J.~S.,  {Kaplinghat} M.,  {Geha} M.,
  {Mu{\~n}oz} R.~R.,  {Simon} J.~D.,   {Avedo} F.~F.,  2010, \mn@doi [MNRAS]
  {10.1111/j.1365-2966.2010.16753.x}, \href
  {http://adsabs.harvard.edu/abs/2010MNRAS.406.1220W} {406, 1220}

\bibitem[\protect\citeauthoryear{{Wu} \& {Kroupa}}{{Wu} \&
  {Kroupa}}{2013a}]{WK13a}
{Wu} X.,  {Kroupa} P.,  2013a, \mn@doi [MNRAS] {10.1093/mnras/stt1332}, \href
  {http://adsabs.harvard.edu/abs/2013MNRAS.435..728W} {435, 728}

\bibitem[\protect\citeauthoryear{{Wu} \& {Kroupa}}{{Wu} \&
  {Kroupa}}{2013b}]{WK13}
{Wu} X.,  {Kroupa} P.,  2013b, \mn@doi [MNRAS] {10.1093/mnras/stt1393}, \href
  {http://adsabs.harvard.edu/abs/2013MNRAS.435.1536W} {435, 1536}

\bibitem[\protect\citeauthoryear{{Wu} \& {Kroupa}}{{Wu} \&
  {Kroupa}}{2015}]{WK15}
{Wu} X.,  {Kroupa} P.,  2015, \mn@doi [MNRAS] {10.1093/mnras/stu2099}, \href
  {http://adsabs.harvard.edu/abs/2015MNRAS.446..330W} {446, 330}

\bibitem[\protect\citeauthoryear{{Yan}, {Jerabkova}  \& {Kroupa}}{{Yan}
  et~al.}{2017}]{Yan+17}
{Yan} Z.,  {Jerabkova} T.,   {Kroupa} P.,  2017, \mn@doi [\aap]
  {10.1051/0004-6361/201730987}, \href
  {http://adsabs.harvard.edu/abs/2017A\%26A...607A.126Y} {607, A126}

\bibitem[\protect\citeauthoryear{{Zhao}}{{Zhao}}{2005}]{Zhao_2005}
{Zhao} H.~S.,  2005, \mn@doi [\aap] {10.1051/0004-6361:200500200}, \href
  {http://adsabs.harvard.edu/abs/2005A\%26A...444L..25Z} {444, L25}

\bibitem[\protect\citeauthoryear{{van Dokkum} \& {Conroy}}{{van Dokkum} \&
  {Conroy}}{2010}]{vanDokkum10}
{van Dokkum} P.~G.,  {Conroy} C.,  2010, \mn@doi [\nat] {10.1038/nature09578},
  \href {http://adsabs.harvard.edu/abs/2010Natur.468..940V} {468, 940}

\bibitem[\protect\citeauthoryear{{van Dokkum} et~al.,}{{van Dokkum}
  et~al.}{2018a}]{vanDokkum+18b}
{van Dokkum} P.,  et~al., 2018a, \mn@doi [Research Notes of the American
  Astronomical Society] {10.3847/2515-5172/aacc6f}, \href
  {http://adsabs.harvard.edu/abs/2018RNAAS...2b..54V} {2, 54}

\bibitem[\protect\citeauthoryear{{van Dokkum} et~al.,}{{van Dokkum}
  et~al.}{2018b}]{vanDokkum+18c}
{van Dokkum} P.,  et~al., 2018b, \mn@doi [\nat] {10.1038/nature25767}, \href
  {http://adsabs.harvard.edu/abs/2018Natur.555..629V} {555, 629}

\bibitem[\protect\citeauthoryear{{van Dokkum} et~al.,}{{van Dokkum}
  et~al.}{2018c}]{vanDokkum+18a}
{van Dokkum} P.,  et~al., 2018c, \mn@doi [ApJl] {10.3847/2041-8213/aab60b},
  \href {http://adsabs.harvard.edu/abs/2018ApJ...856L..30V} {856, L30}

\bibitem[\protect\citeauthoryear{{van Dokkum}, {Danieli}, {Romanowsky},
  {Abraham}  \& {Conroy}}{{van Dokkum} et~al.}{2019a}]{vD19}
{van Dokkum} P.,  {Danieli} S.,  {Romanowsky} A.,  {Abraham} R.,   {Conroy} C.,
   2019a, \mn@doi [Research Notes of the American Astronomical Society]
  {10.3847/2515-5172/ab05d6}, \href
  {http://adsabs.harvard.edu/abs/2019RNAAS...3b..29V} {3, 29}

\bibitem[\protect\citeauthoryear{{van Dokkum}, {Danieli}, {Abraham}, {Conroy}
  \& {Romanowsky}}{{van Dokkum} et~al.}{2019b}]{vD19_DF4}
{van Dokkum} P.,  {Danieli} S.,  {Abraham} R.,  {Conroy} C.,   {Romanowsky}
  A.~J.,  2019b, \mn@doi [ApJL] {10.3847/2041-8213/ab0d92}, \href
  {https://ui.adsabs.harvard.edu/#abs/2019arXiv190105973V} {874, L5}

\makeatother
\end{thebibliography}

\appendix \section{The velocity dispersion of NGC 1052-DF2: Is a Gaussian description valid?}
\label{Appendix}

The original analysis of DF2's internal kinematics \citep{vanDokkum+18c} was based on only ten GCs. It has been shown that such a small sample can lead to significant uncertainties in estimates of the inferred velocity dispersion \citep{Laporte_2019}. In Section \ref{sec:veldisp}, we therefore applied a standard Gaussian analysis to the ten reported data points, including the updated value from \citet{vanDokkum+18b}. There, we averaged the error budgets given by \citet{vanDokkum+18c} to obtain the uncertainty of each measurement. For example, GC 59 with radial velocity of $1799^{+16}_{-15}$ km/s was assumed to have a radial velocity of $1799 \pm 15.5$ km/s.

One possible objection to our method is that some subsample of the data is unusually clustered in velocity space (i.e. unusually dynamically cold). However, we must bear in mind that when considering over 1000 possible ways of obtaining a subsample from the ten observed GCs, some combinations are bound to yield a much smaller velocity dispersion than that of the underlying population. To illustrate this point, we note that \citet{vanDokkum+18c} argued against a Gaussian distribution of width 10 km/s (near our most likely value) because 6/10 GCs have a radial velocity within ${\pm 4}$ km/s of their mean. However, those authors did not mention that there are 210 ways of choosing 6 objects out of 10.

Before considering this in detail, a rough calculation illustrates why such a clustering of radial velocities is not very unlikely. For a Gaussian distribution, the probability of lying within $\frac{4}{10}\sigma$ of the mean is 0.31. For this to happen with 6/10 objects has a chance of 0.04 using standard binomial statistics. However, the actual probability is even higher because we have assumed that the aforementioned 6/10 objects have radial velocities within ${\pm 4}$ km/s of the true mean, when what is observed is that they are simply within an 8 km/s range of each other.

To more rigorously check whether the observed radial velocities are consistent with a Gaussian distribution, we conduct $10^6$ Monte Carlo trials in which each radial velocity follows a normal distribution of width $\sigma_i$ (Equation \ref{sigma_total}). We then find all subsamples of size $n \geq 2$. For each $n$, we find the proportion of mock datasets which have a subsample of size $n$ that is dynamically colder than the coldest observed subsample of that size. Our results are shown in Figure \ref{NGC_1052_subsamples} for three choices of $\sigma_{int}$.

At the 99\% confidence level, all subsample velocity dispersions are consistent with our model for ${\sigma = 10}$ or 15 km/s (Figure \ref{NGC_1052_subsamples}). The fact that this is true for ${n = 9}$ indicates that no single object is a statistically significant outlier, casting doubt on the contrary claim of \citet{vanDokkum+18c}. Although some tension is apparent for the ${n = 6}$ case when $\sigma = 20$ km/s, it must be borne in mind that there is no a priori reason to consider the coldest 6/10 subsample rather than e.g. the coldest 5/10 or 7/10. Given that over 1000 distinct subsamples can be drawn from 10 objects but only 210 of these subsamples have size ${n = 6}$, it is clear that there are significant `look elsewhere' effects which we do not take into account. Thus, the probabilities shown in Figure \ref{NGC_1052_subsamples} must be considered underestimates. Even if they are taken at face value, it is clear that there is no compelling reason to reject our \emph{standard} assumptions, at least if $\sigma_{int}$ is within the 68\% confidence interval suggested by our analysis (Figure \ref{NGC_1052_marginal_sigma}). Thus, an intrinsic velocity dispersion of 14 km/s is entirely consistent with all available observations and also with MOND, as demonstrated in this work using analytic and numerical methods.

For future reference, Table \ref{Confidence_levels_sigma} shows the 1, 2 and 3 sigma confidence intervals of the velocity dispersion as derived here from the 10 GC data points in \citet{vanDokkum+18b}, updating the data in \citet{vanDokkum+18c}.

%In fact, it is difficult to see how a much smaller dispersion would be consistent with the very low radial velocity of globular cluster 98. However, this object is unlikely to be a chance superposition given its proximity to DF2 on the sky \citep[][figure 1]{van_Dokkum_2018} and the fact that its radial velocity differs only $\approx 40$ km/s from the mean of the other nine globular clusters.

\begin{figure}
	\centering
		\includegraphics[width = 8.4cm] {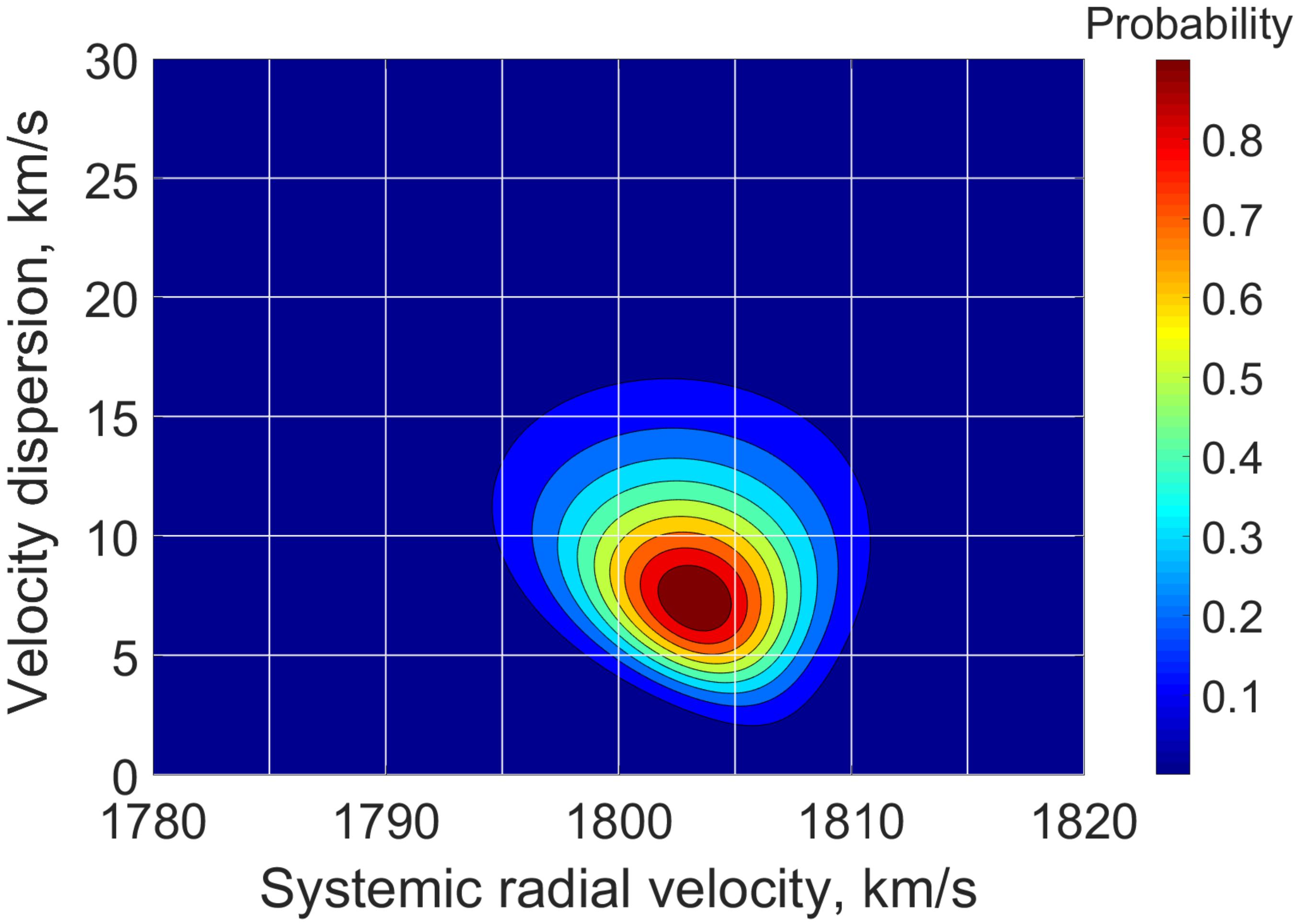}
		\caption{The probabilities of different models relative to the most likely model (systemic velocity of 1802 km/s, velocity dispersion of 8.0 km/s).}
	\label{NGC_1052_contour}
\end{figure}

\begin{figure}
	\centering
		\includegraphics[width = 8.4cm] {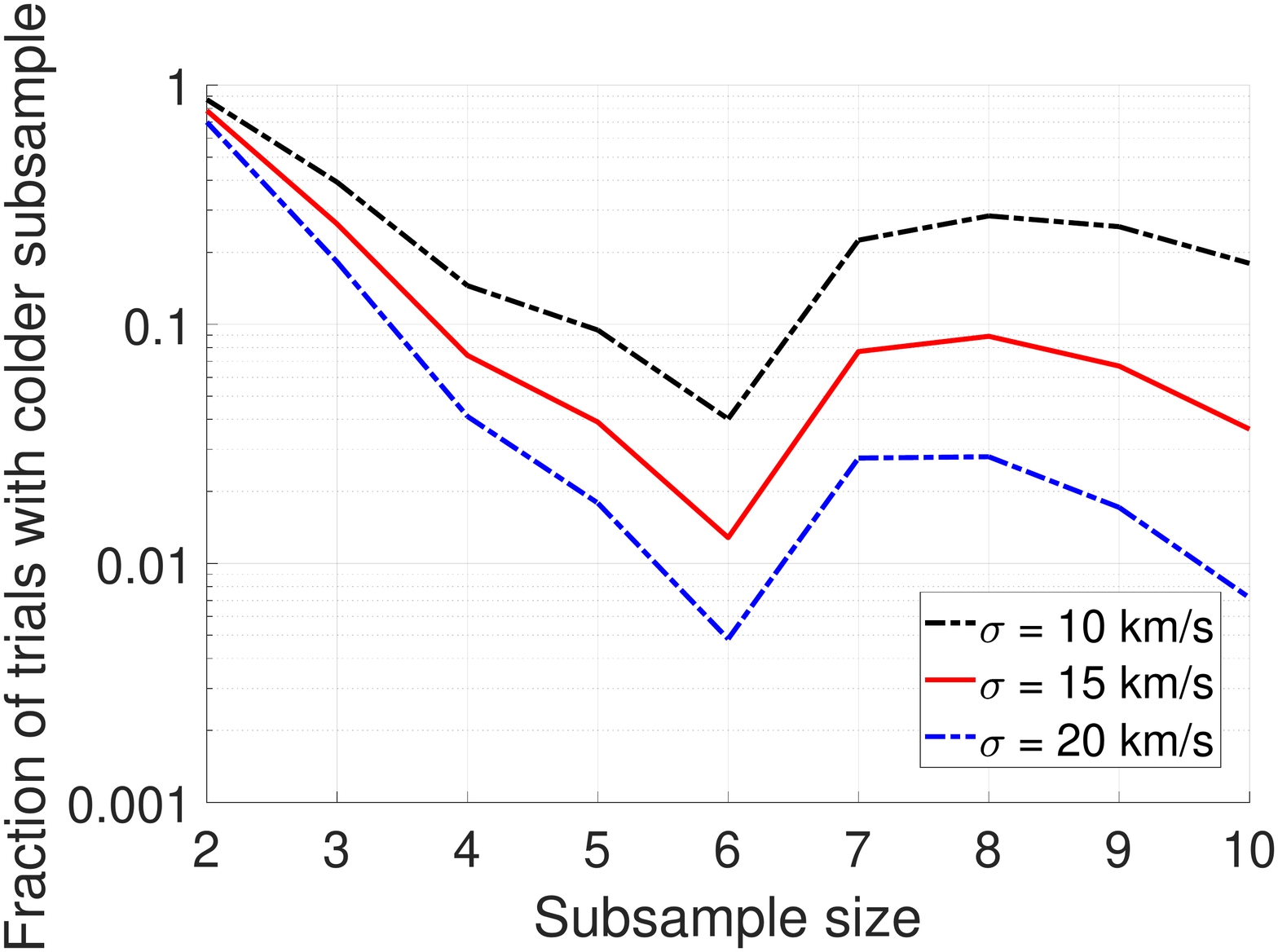}
		\caption{The proportion of $10^6$ Monte Carlo trials in which the mock data has a subsample that is dynamically colder than the coldest observed subsample of the indicated size, shown for different subsample sizes. We show results for an intrinsic velocity dispersion of $\sigma_{int} = 10$ km/s (dash-dotted black), 15 km/s (solid red) and 20 km/s (dashed blue), requiring us to construct ${3 \times 10^6}$ mock datasets altogether. Although mild tension is apparent for the 6/10 case (especially for $\sigma_{int} = 20$ km/s), there is no a priori reason to suspect that 6 of the 10 GCs might have an unusually low velocity dispersion rather than e.g. 5/10 or 8/10. This significant `look elsewhere' effect is not taken into account here, so these probabilities should be treated as underestimates.}
	\label{NGC_1052_subsamples}
\end{figure}

%From Oliver on the extent on the sky of the system which may affect the velocities due to slighlty different viewing angles from one side to the other.
%I think what Joerg means is that the motion of the Sun will induce a velocity gradient in the sky. One which needs to be corrected for in %extended systems. Here you can read this up (https://advlabwiki.johnshopkins.edu/images/Vlsr.pdf).
%In short it is this formula:
%vR=20*(cos(270)*cos(30)*cos(ra)*cos(dec)+sin(270)*cos(30)*sin(ra)*cos(dec)+sin(30)*sin(dec))
%where (270,30) are the coordinates the Sun is heading to (near Vega) and (ra,dec) are the coordinates of an object in the sky.
%I checked what the differences are for different positions of the GCs in DF2, and it is in the order of 0.002 km/s. So this will not affect the analysis. Which is also what I expected because the extend is too tiny in the sky. Even the Centaurus A dwarf system spread over several %degrees is affected by only +- 1 km/s due to this.
%If this is not what Joerg means, than at least I checked that the Sun's motion does not induce a velocity gradient. :-)
\label{lastpage}
\end{document}